\documentclass[longauth]{aa}  

\usepackage{graphicx}
\usepackage{txfonts}
\usepackage{amsmath}	
\usepackage[usenames,dvipsnames]{color}

\newcommand{\dd}{\mathrm{d}}

\begin{document}

   \title{Constraints on primordial non-Gaussianity from the cross-correlation of DESI luminous red galaxies and $Planck$ CMB lensing}

\authorrunning{J.~R.~Bermejo-Climent et al.}

\author{
J.~R.~Bermejo-Climent \inst{1,2,3}\thanks{\email{jrbermejo@iac.es}},
R.~Demina \inst{2},
A.~Krolewski \inst{4,5,6},
E.~Chaussidon \inst{7},
M.~Rezaie \inst{8},
S.~Ahlen \inst{9},
S.~Bailey \inst{7},
D.~Bianchi \inst{10},
D.~Brooks \inst{11},
E.~Burtin \inst{12},
T.~Claybaugh \inst{7},
A.~de la Macorra \inst{13},
Arjun~Dey \inst{14},
P.~Doel \inst{11},
G.~Farren \inst{7},
S.~Ferraro \inst{7,15},
J.~E.~Forero-Romero \inst{16,17},
E.~Gaztañaga \inst{18,19,20},
S.~Gontcho A Gontcho \inst{7},
G.~Gutierrez \inst{21},
C.~Hahn \inst{22},
K.~Honscheid \inst{23,24,25},
C.~Howlett \inst{26},
R.~Kehoe \inst{27},
D.~Kirkby \inst{28},
T.~Kisner \inst{7},
M.~Landriau \inst{7},
L.~Le~Guillou \inst{29},
M.~E.~Levi \inst{7},
M.~Manera \inst{30,31},
A.~Meisner \inst{14},
R.~Miquel \inst{32,31},
J.~Moustakas \inst{33},
J.~ A.~Newman \inst{34},
G.~Niz \inst{35,36},
N.~Palanque-Delabrouille \inst{12,7},
W.~J.~Percival \inst{4,5,6},
F.~Prada \inst{37},
I.~P\'erez-R\`afols \inst{38},
D.~Rabinowitz \inst{39},
A.~J.~Ross \inst{23,40,25},
G.~Rossi \inst{41},
E.~Sanchez \inst{42},
D.~Schlegel \inst{7},
D.~Sprayberry \inst{14},
G.~Tarl\'{e} \inst{43},
B.~A.~Weaver \inst{14},
M.~White \inst{44,15},
C.~Yèche \inst{12},
and P.~Zarrouk \inst{29},
}

\institute{
Instituto de Astrof\'{\i}sica de Canarias, C/ V\'{\i}a L\'{a}ctea, s/n, E-38205 La Laguna, Tenerife, Spain 
\and Department of Physics \& Astronomy, University of Rochester, 206 Bausch and Lomb Hall, P.O. Box 270171, Rochester, NY 14627-0171, USA 
\and Departamento de Astro\'{\i}sica, Universidad de La Laguna, Avenida Francisco S\'{a}nchez, s/n, E-38205 La Laguna, Tenerife, Spain 
\and Department of Physics and Astronomy, University of Waterloo, 200 University Ave W, Waterloo, ON N2L 3G1, Canada 
\and Perimeter Institute for Theoretical Physics, 31 Caroline St. North, Waterloo, ON N2L 2Y5, Canada 
\and Waterloo Centre for Astrophysics, University of Waterloo, 200 University Ave W, Waterloo, ON N2L 3G1, Canada 
\and Lawrence Berkeley National Laboratory, 1 Cyclotron Road, Berkeley, CA 94720, USA 
\and Department of Physics, Kansas State University, 116 Cardwell Hall, Manhattan, KS 66506, USA 
\and Physics Dept., Boston University, 590 Commonwealth Avenue, Boston, MA 02215, USA 
\and Dipartimento di Fisica ``Aldo Pontremoli'', Universit\`a degli Studi di Milano, Via Celoria 16, I-20133 Milano, Italy 
\and Department of Physics \& Astronomy, University College London, Gower Street, London, WC1E 6BT, UK 
\and IRFU, CEA, Universit\'{e} Paris-Saclay, F-91191 Gif-sur-Yvette, France 
\and Instituto de F\'{\i}sica, Universidad Nacional Aut\'{o}noma de M\'{e}xico,  Circuito de la Investigaci\'{o}n Cient\'{\i}fica, Ciudad Universitaria, Cd. de M\'{e}xico  C.~P.~04510,  M\'{e}xico 
\and NSF NOIRLab, 950 N. Cherry Ave., Tucson, AZ 85719, USA 
\and University of California, Berkeley, 110 Sproul Hall \#5800 Berkeley, CA 94720, USA 
\and Departamento de F\'isica, Universidad de los Andes, Cra. 1 No. 18A-10, Edificio Ip, CP 111711, Bogot\'a, Colombia 
\and Observatorio Astron\'omico, Universidad de los Andes, Cra. 1 No. 18A-10, Edificio H, CP 111711 Bogot\'a, Colombia 
\and Institut d'Estudis Espacials de Catalunya (IEEC), c/ Esteve Terradas 1, Edifici RDIT, Campus PMT-UPC, 08860 Castelldefels, Spain 
\and Institute of Cosmology and Gravitation, University of Portsmouth, Dennis Sciama Building, Portsmouth, PO1 3FX, UK 
\and Institute of Space Sciences, ICE-CSIC, Campus UAB, Carrer de Can Magrans s/n, 08913 Bellaterra, Barcelona, Spain 
\and Fermi National Accelerator Laboratory, PO Box 500, Batavia, IL 60510, USA 
\and Department of Astrophysical Sciences, Princeton University, Princeton NJ 08544, USA 
\and Center for Cosmology and AstroParticle Physics, The Ohio State University, 191 West Woodruff Avenue, Columbus, OH 43210, USA 
\and Department of Physics, The Ohio State University, 191 West Woodruff Avenue, Columbus, OH 43210, USA 
\and The Ohio State University, Columbus, 43210 OH, USA 
\and School of Mathematics and Physics, University of Queensland, Brisbane, QLD 4072, Australia 
\and Department of Physics, Southern Methodist University, 3215 Daniel Avenue, Dallas, TX 75275, USA 
\and Department of Physics and Astronomy, University of California, Irvine, 92697, USA 
\and Sorbonne Universit\'{e}, CNRS/IN2P3, Laboratoire de Physique Nucl\'{e}aire et de Hautes Energies (LPNHE), FR-75005 Paris, France 
\and Departament de F\'{i}sica, Serra H\'{u}nter, Universitat Aut\`{o}noma de Barcelona, 08193 Bellaterra (Barcelona), Spain 
\and Institut de F\'{i}sica d’Altes Energies (IFAE), The Barcelona Institute of Science and Technology, Edifici Cn, Campus UAB, 08193, Bellaterra (Barcelona), Spain 
\and Instituci\'{o} Catalana de Recerca i Estudis Avan\c{c}ats, Passeig de Llu\'{\i}s Companys, 23, 08010 Barcelona, Spain 
\and Department of Physics and Astronomy, Siena College, 515 Loudon Road, Loudonville, NY 12211, USA 
\and Department of Physics \& Astronomy and Pittsburgh Particle Physics, Astrophysics, and Cosmology Center (PITT PACC), University of Pittsburgh, 3941 O'Hara Street, Pittsburgh, PA 15260, USA 
\and Departamento de F\'{\i}sica, DCI-Campus Le\'{o}n, Universidad de Guanajuato, Loma del Bosque 103, Le\'{o}n, Guanajuato C.~P.~37150, M\'{e}xico. 
\and Instituto Avanzado de Cosmolog\'{\i}a A.~C., San Marcos 11 - Atenas 202. Magdalena Contreras. Ciudad de M\'{e}xico C.~P.~10720, M\'{e}xico 
\and Instituto de Astrof\'{i}sica de Andaluc\'{i}a (CSIC), Glorieta de la Astronom\'{i}a, s/n, E-18008 Granada, Spain 
\and Departament de F\'isica, EEBE, Universitat Polit\`ecnica de Catalunya, c/Eduard Maristany 10, 08930 Barcelona, Spain 
\and Physics Department, Yale University, P.O. Box 208120, New Haven, CT 06511, USA 
\and Department of Astronomy, The Ohio State University, 4055 McPherson Laboratory, 140 W 18th Avenue, Columbus, OH 43210, USA 
\and Department of Physics and Astronomy, Sejong University, 209 Neungdong-ro, Gwangjin-gu, Seoul 05006, Republic of Korea 
\and CIEMAT, Avenida Complutense 40, E-28040 Madrid, Spain 
\and University of Michigan, 500 S. State Street, Ann Arbor, MI 48109, USA 
\and Department of Physics, University of California, Berkeley, 366 LeConte Hall MC 7300, Berkeley, CA 94720-7300, USA 
}

   \date{Received XXX; accepted YYY}

  \abstract
   {}
   {We use the angular cross-correlation between a luminous red galaxy (LRG) sample from the  Dark Energy Spectroscopic Instrument (DESI) Legacy Survey data release DR9 and the $Planck$ cosmic microwave background (CMB) lensing maps to constrain the local primordial non-Gaussianity parameter, $f_{\rm NL}$, using the scale-dependent galaxy bias effect. The galaxy sample covers approximately 40\% of the sky, contains galaxies up to redshift $z \sim 1.4$, and is calibrated with the LRG spectra that have been observed for DESI Year 1 (Y1).}
   {We apply a nonlinear imaging systematics treatment based on neural networks to remove observational effects that could potentially bias the $f_{\rm NL}$ measurement. Our measurement is performed without blinding, but the full analysis pipeline is tested with simulations including systematics.}
   {Using the two-point angular cross-correlation between LRG and CMB lensing only, we find $f_{\rm NL} = 39_{-38}^{+40}$ at the 68\% confidence level, and our result is robust in terms of systematics and cosmological assumptions. If we combine this information with the autocorrelation of LRG, applying a scale cut to limit the impact of systematics, we find $f_{\rm NL} = 24_{-21}^{+20}$ at the 68\% confidence level. Our results motivate the use of CMB lensing cross-correlations to measure $f_{\rm NL}$ with future datasets, given its stability in terms of observational systematics compared to the angular autocorrelation. Furthermore, performing accurate systematics mitigation is crucially important in order to achieve competitive constraints on $f_{\rm NL}$ from CMB lensing cross-correlation in combination with the tracers' autocorrelation.}
{}


\maketitle

\section{Introduction}
Cosmic inflation was proposed as a theory in the early 1980s \citep{Guth:1980zm,STAROBINSKY198099}. The inflation framework was initially formulated to solve big bang problems such as the horizon, flatness, and magnetic monopole problems; however, inflation is also able to explain the formation of primordial density perturbations \citep{STAROBINSKY1982175,Guth:1985ya,Baarden}. Inflation is defined as a phase in which the Universe is expanding exponentially, driven by a scalar field, $\phi$. Several models of inflation have been proposed in the literature (see e.g., \citealt{Langlois_2010,Vazquez_Gonzalez_2020} for a review). The model of inflation and its predictions are defined by choosing the form of the potential, $V(\phi)$. The simplest inflationary models predict Gaussian initial conditions; however, alternative inflationary models predict different levels of non-Gaussianity in the primordial density perturbations \citep{Chen_2010,10.1093/ptep/ptu060}. The level of non-Gaussianity has usually been characterized in the literature with the $f_{\rm NL}$ non-Gaussianity parameter, such that detecting $f_{\rm NL} \neq 0$ is a signature of having non-Gaussian initial conditions.

The tightest constraint on $f_{\rm NL}$ is currently provided by the measurements from the cosmic microwave background (CMB) bispectrum. Using $Planck$ 2018 data, a value $f_{\rm NL} = -0.9 \pm 5.1$ at the 68\% confidence level is found \citep{planckcollaboration2019planck}. However, \cite{PhysRevD.77.123514} first noticed that local non-Gaussian initial conditions lead to a characteristic scale-dependent signature in the galaxy bias, following a $1/k^2$ scale-dependence in the ratio between the total matter to observed density of galaxies. During the last decade, many works have performed measurements from the large-scale structure using the scale-dependent bias effect (\citealt{Ross_2012,Castorina_2019,muller,cabass,damico2023limits}, among others). In the last years, some measurements from large scale structure (LSS) using the scale-dependent galaxy bias have been achieved using the quasars from 
the data release DR16 of the Extended Baryon Oscillation Spectroscopic Survey (eBOSS): \cite{muller} measure $f_{\rm NL} = -12 \pm 21$ and \cite{Cagliari_2024} obtain $-4 < f_{\rm NL} < 27$ using different methodologies. More recently, \cite{chaussidon2024constrainingprimordialnongaussianitydesi} have improved this constraint to $f_{\rm NL} = -3.6^{+9.1}_{-9.0}$ using the 3D power spectrum of the Dark Energy Spectroscopic Instrument (DESI) data release DR1 galaxies and quasars. This is a challenging measurement because it requires very accurate control of the largest scales where the scale-dependent bias effect due to $f_{\rm NL}$ arises. Further, it is also important to mention that, when performing $f_{\rm NL}$ measurements from LSS, unless using certain assumptions we actually measure the product of $f_{\rm NL}$ times an unknown bias (see Sect.~\ref{sec:png} for more details).

DESI \citep{Snowmass2013.Levi} is a spectroscopic survey that is currently being carried out from the 4m Mayall telescope at Kitt Peak National Observatory (Arizona, USA). Its unique design with 5000 fibers with robotic positioners allows it to take thousands of spectra in a single exposure \citep{DESI2016b.Instr,DESI2022.KP1.Instr,FocalPlane.Silber.2023,Corrector.Miller.2023,Spectro.Pipeline.Guy.2023,SurveyOps.Schlafly.2023}. Theoretical forecasts \citep{DESI2016a.Science} expect that the full five-year survey will have the ability to achieve a sensitivity $\sigma(f_{\rm NL}) \sim 5$, similar to the best current CMB bispectrum constraint, if there is a good control of systematic effects. Before the first spectroscopic data releases \citep{DESI2023b.KP1.EDR,DESI2023a.KP1.SV} and science results \citep{2024arXiv240403000D,2024arXiv240403001D,2024arXiv240403002D,desicollaboration2024desi2024vfullshape,desicollaboration2024desi2024viicosmological} came out, a full imaging survey was performed in order to select the spectroscopic targets. This targeting survey is called the DESI Legacy Survey \citep{1804.08657} and covers a broad area ($\gtrsim$20000 deg$^2$), making it useful for measuring $f_{\rm NL}$ using the scale-dependent galaxy bias. Two previous works have already used the DESI Legacy Survey information to put a constraint on $f_{\rm NL}$: \cite{rezaie2023local} used the angular power spectra of the louminous red galaxy (LRG) targets, and \cite{Krolewski_2024} used the cross-correlation between quasar targets and the $Planck$ CMB lensing.

CMB lensing describes the remapping of the CMB anisotropies due to gravitational lensing by structures along the line of slight. The CMB lensing potential can be easily measured from the observations of the lensed sky \citep{Hu_2002} and was first detected by \cite{Smith_2007}. Since it contains information about the large-scale structure geometry, its cross-correlation with galaxy tracers can be useful to constrain cosmology. Although CMB lensing and galaxy tracers probe the same structures, they are affected by different systematics, making the cross-correlation between the two a powerful additional tool for measurements into the systematics-dominated regime.
Several authors have stressed, using theoretical forecasts, the capabilities of the cross-correlation between CMB lensing and galaxy matter tracers to better constrain $f_{\rm NL}$ (e.g., \citealt{PhysRevD.80.123527,Schmittfull_2018,Giusarma_2018,Ballardini_2019,Bermejo_Climent_2021,Perna:2023dgg}). More recently, \cite{Krolewski_2024} find $f_{\rm NL} = -26^{+45}_{-40}$ using the cross-correlation between $Planck$ lensing and DESI quasar targets. Additionally, recent data analysis works have been performed to constrain other cosmological parameters such as the amplitude of matter density fluctuations, commonly parametrized in terms of $\sigma_8$ (the RMS density contrast smoothed on a scale of 8 h/Mpc), and matter density, $\Omega_{\rm m}$, using CMB cross-correlations with the DESI Legacy Survey (e.g., \citealt{White_2022}, \citealt{sailer2024cosmologicalconstraintscrosscorrelationdesi,kim2024atacamacosmologytelescopedr6}).

In this paper, we intend to extend the analysis done by \cite{rezaie2023local} with the DESI LRG sample to the inclusion of the CMB lensing cross-correlation as additional information and an additional technique to limit the impact of observational systematics. In \cite{rezaie2023local}, an extensive and detailed effort was made to remove observational systematics that could bias the primordial non-Gaussianity (PNG) measurement. Nonetheless, they conclude that their results motivate further studies of PNG with samples less sensitive to systematics like LRG spectroscopic data. Here, our aim is to explore the ability of CMB lensing - LRG cross-correlation to constrain $f_{\rm NL}$ and its stability in terms of systematics, alone and in combination with the LRG autocorrelation.

In Sect.~\ref{sec:theory} we review the theoretical framework for the imprints of $f_{\rm NL}$ in a scale-dependent galaxy bias and for the cosmological observables we study in the angular domain. In Sect.~\ref{sec:data} we present the DESI LRG and $Planck$ lensing datasets used for our analysis. In Sect.~\ref{sec:pipeline} we describe the methodology followed for treating our datasets, including a systematics mitigation, computation of observables, and parameter inference pipeline. In Sect.~\ref{sec:mocks} we discuss a validation of our analysis pipeline with mock LRG and CMB lensing fields. In Sect.~\ref{sec:results} we present our results for the measurement of PNG and some robustness tests, and in Sect.~\ref{sec:conclusions} we summarize our conclusions.

\section{Theory}
\label{sec:theory}
In this section we first provide a description of the physical model that originates a scale-dependent galaxy bias due to a local PNG. We then describe the cosmological observables in the 2D harmonic space involved in our analysis.

\subsection{Primordial non-Gaussianity and scale-dependent bias}
\label{sec:png}
If we assume a type of non-Gaussianity that depends only on the local value of the potential, the parametrization of the primordial potential can be written as \citep{PhysRevD.63.063002}
\begin{equation}
    \Phi = \phi + f_{\rm NL} ( \phi^2 - \langle \phi \rangle ^2) \,,
\end{equation}
where $f_{\rm NL}$ is the parameter that describes the amplitude of the quadratic non-Gaussian term and $\phi$ is a random Gaussian field. We studied $f_{\rm NL}$ through its impact on the scale-dependent galaxy bias, as introduced in \citet{PhysRevD.77.123514}. If we assume the so-called universality relation \citep{Slosar_2008}, the contribution to the galaxy bias is expressed as
\begin{equation}
\label{deltab}
\Delta b(k,z) = 2(b_g-p) f_{\rm 
NL}\frac{\delta_{\rm crit}}{\alpha(k)} \,,
\end{equation}
where $\delta_{\rm crit}$ = 1.686 is the threshold overdensity for spherical collapse, $b_g$ is the galaxy bias as a function of the redshift $z$, $p$ is a parameter characterizing the tracers' response to PNG, assumed as $p \simeq 1$ for the case of LRG, and $\alpha(k)$ is the relation between potential and density field, such that $\delta(k) = \alpha(k) \Phi(k)$. The value of $\alpha (k)$ is given by
\begin{equation}
    \alpha(k) = \frac{2 k^2 T(k) D(z)}{3 \Omega_{\rm m}} \frac{c^2}{H_0^2} \frac{g(0)}{g (\infty)} \,,
\end{equation}
where $T(k)$ is the transfer function, $D(z)$ is the growth factor (normalized to be 1 at $z=0$), $\Omega_{\rm m}$ the matter density, and the factor $g(\infty)/g(0) \simeq 1.3$ accounts for the different normalizations of $D(z)$ in the CMB and LSS literature. This definition of $f_{\rm NL}$ is therefore the so-called CMB convention. Other authors (e.g., \citealt{Carbone_2008,Afshordi_2008,Grossi_2009}) refer to the use of the `LSS convention,' where the $g(\infty)/g(0)$ factor is absorbed into the definition of  $f_{\rm NL}$, such that $f_{\rm NL}^{\rm LSS} \simeq 1.3 f_{\rm NL}^{\rm CMB}$. Throughout this paper we use $p = 1$ as a baseline; however, many works based on dark-matter-only simulations (e.g., \citealt{adame}) find significant deviations from $p=1$. In this direction, other authors (e.g., \citealt{Barreira_2020,Barreira_2022}) have stressed that we can only constrain the product $b_\phi f_{\rm NL}$ through the scale-dependent bias effect, where  $b_\phi$ is  a parameter usually defined as $b_\phi = 2 \delta_{\rm crit} (b_g - p)$, in order to account for the uncertainties on $p$. 

\subsection{Cosmological observables}

In this work, we focus on the angular power spectrum of the galaxy - CMB lensing cross-correlation, $C_\ell ^{\kappa G}$, as well as the galaxy autocorrelation, $C_\ell ^{GG}$. The angular power spectrum can be calculated as
\begin{equation} \label{eqn:APS}
    C_\ell^{XY} = 4\pi \int \frac{\dd k}{k} {\cal P}(k) I^X_{\ell}(k) I^Y_{\ell}(k),
\end{equation}
where ${\cal P}(k) \equiv k^3 P(k) / (2\pi^2)$ is the dimensionless primordial power spectrum 
and $I^X_{\ell}(k)$ is the kernel for the $X$ field for unit of primordial power spectrum.

All the weak lensing quantities can be defined from the lensing potential
\begin{equation}
    \phi\left(\hat{\bf n},\chi\right) = \frac{2}{c^2} \int_0^\chi \dd \chi' \frac{\chi-\chi'}{\chi\chi'}
    \Phi\left(\chi'\hat{\bf n},\chi'\right),
\end{equation}
where $\Phi\left(\hat{\bf n},\chi\right)$ is the gravitational potential. The comoving distance is
\begin{equation}
    \chi(z) = \int_0^z \frac{c\, \dd z'}{H(z')}\,. 
\end{equation}
The observable two-dimensional lensing potential, averaged over background sources with a redshift 
distribution $W_b\left(\chi\right)$, 
is given by
\begin{equation}
    \phi\left(\hat{\bf n}\right) = \frac{2}{c^2} \int_0^\chi \frac{\dd \chi'}{\chi'} q_b\left(\chi'\right)
    \Phi\left(\chi'\hat{\bf n},\chi'\right),
\end{equation}
where $q_b\left(\chi\right)$ is the lensing efficiency for a given background distribution, $W_b$, 
defined as
\begin{equation}
    q_b\left(\chi\right) = \int_0^\chi \dd \chi' \frac{\chi'-\chi}{\chi'} W_b\left(\chi'\right) \,.
\end{equation}

By expanding the gravitational potential in Fourier space and using the plane-wave expansion, we can define the lensing potential kernel as
\begin{equation} \label{eqn:kernel_phi}
    I^\phi_\ell (k) = 2\left(\frac{3\Omega_mH_0^2}{2k^2c^2}\right) \int \frac{\dd \chi}{(2\pi)^{3/2}}
    \frac{q_b\left(\chi\right)}{\chi a(\chi)} j_\ell\left(k\chi\right) 
    \delta\left(k,\chi\right) \,,
\end{equation}
where $\Omega_m$ is the present-day matter density, $a(\chi)$ is the scale factor, $H_0$ is the Hubble constant, $\delta(k,\chi)$ 
is the comoving-gauge linear matter density perturbation, and $j_\ell$ the spherical Bessel functions. 
In case of CMB lensing, the source distribution can be approximated by 
$W_{\rm CMB} \left(\chi\right) \simeq \delta_{\rm D}\left(\chi-\chi_*\right)$ and the lensing efficiency by
\begin{equation}
    q_{\rm CMB}\left(\chi\right) \simeq \frac{\chi_*-\chi}{\chi_*},
\end{equation}
where $\chi_*$ is the comoving distance at the surface of last scattering, and Eq.~\eqref{eqn:kernel_phi} 
reduces to
\begin{equation}
 \label{eqn:kernel_CMB}
    I^{\phi_{\rm CMB}}_\ell (k) = 2\left(\frac{3\Omega_mH_0^2}{2k^2c^2}\right) \int \frac{\dd \chi}{(2\pi)^{3/2}} \frac{\chi_*-\chi}{\chi_*\chi} \frac{1}{a(\chi)}
    j_\ell\left(k\chi\right) \delta\left(k,\chi\right) \,.
\end{equation}
Finally, the convergence $\kappa = \nabla^2 \phi/2$ can be expanded in spherical harmonics as
\begin{equation}
    \kappa\left({\bf \hat{n}}\right) = -\frac{1}{2} \sum_{\ell, m} \ell(\ell+1)\phi_{\ell m}Y_\ell^{m}\left({\bf \hat{n}}\right) \,.
\end{equation}
We then relate the two kernels by
\begin{equation} \label{eqn:kernel_kappa}
    I_\ell^{\kappa}(k) = \frac{\ell(\ell+1)}{2} I_\ell^{\phi}(k) \,.
\end{equation}

The two-dimensional integrated window function for the galaxy number counts is
\begin{equation} \label{eqn:kernel_counts}
    I^G_\ell(k) = \int \frac{\dd \chi}{(2\pi)^{3/2}} W(\chi) \Delta_{\ell}(k,\chi), 
\end{equation}
where $\Delta_{\ell}(k,\chi)$ is the observed number count and $W(\chi)$ is a window function given by the redshift distribution of galaxies. At first order, the most important contribution to  $\Delta_{\ell}(k,\chi)$ is given by the synchronous gauge source counts Fourier transformed and expanded into multipoles, 
$\Delta^s_{\ell}(k,\chi)$.  We assumed that $\Delta^s_{\ell}(k,\chi)$ is related to the underlying matter density field through a scale- and redshift-dependent galaxy bias, $b_g$, as 
\begin{equation}
    \Delta^s_{\ell}(k,\chi) = b_g(k,\chi) \delta(k,\chi) j_\ell\left(k\chi\right),
\end{equation}
where $b_g(k,\chi)$ is given by the sum of a linear bias, which is not scale dependent, plus the scale-dependent contribution given by Eq.~\ref{deltab}. We also considered nonlinear contributions to the power spectrum using \texttt{halofit} \citep{Takahashi_2012}.
In this paper, we also consider two important contributions to the total observed number counts: the effects of redshift space dirtortions (RSD) and lensing magnification (see Fig.~\ref{fig:sz} for more details). The RSD term is given by
\begin{equation}
    \Delta^{\rm RSD}_\ell (k,\chi) = \frac{k v_k}{\cal H} j_\ell''(k\chi), 
\end{equation}
where $v_k$ is the velocity of the sources and ${\cal H}$ is the conformal Hubble parameter. The lensing convergence contribution is given by
\begin{multline}
\label{eq:deltalens}
\Delta^{\rm lensing}_\ell (k,\chi) =\ell(\ell+1) (2-5s)
\int_0^\chi {\rm d}\chi' \frac{\chi-\chi'}{\chi\chi'}
\Phi\left(\chi'\hat{\bf n},\chi'\right),
\end{multline}
where $s$ is the magnification bias that accounts for the fact that observed galaxies are magnified by gravitational lensing. We note that there are other contributions from general relativity to the number counts, but we consider them of second order since the most important contribution to $C_\ell^{\kappa G}$ is the lensing magnification (see e.g., Appendix A of \citealt{Bermejo_Climent_2021}). 

In practice, to numerically evaluate the observables described in this section, we used the Limber approximation \citep{limber} for $\ell \geq 100$ and computed the full quantities for $\ell < 100$,
following the implementation in the \texttt{CAMBSOURCES} module \citep{Challinor:2011bk}.
This was done in order to improve the computational efficiency while having a good estimation of the theory model at the lower multipoles, where the Limber approximation is less accurate. More details on the code used for the computation of theoretical observables can be found in Sect.~\ref{sec:likelihood}.

\section{Datasets}
\label{sec:data}
We describe in this section the datasets involved in our analysis. The two main ingredients were an LRG photometric catalog from the DR9 Legacy Survey \citep{Zhou_2023b} and the $Planck$ PR4 public CMB lensing maps \citep{Carron_2022}. We also used an LRG spectroscopic sample from the DESI Survey Validation data to calibrate the redshift distribution of the photometric DESI LRGs.
\subsection{Luminous red galaxies}

\begin{figure}
    \centering
\includegraphics[width=0.5\textwidth]{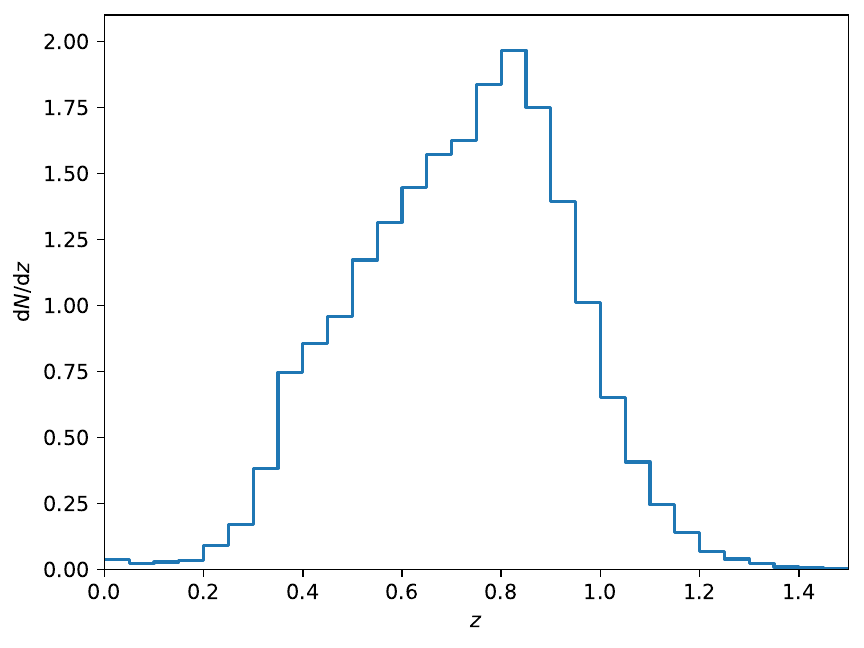}
    \caption{Normalized redshift distribution of the LRG sample, directly measured using the spectroscopic redshifts from DESI Y1 data.}
    \label{fig:dndz}
\end{figure}

\begin{figure*}
    \centering
\includegraphics[width=0.49\textwidth]{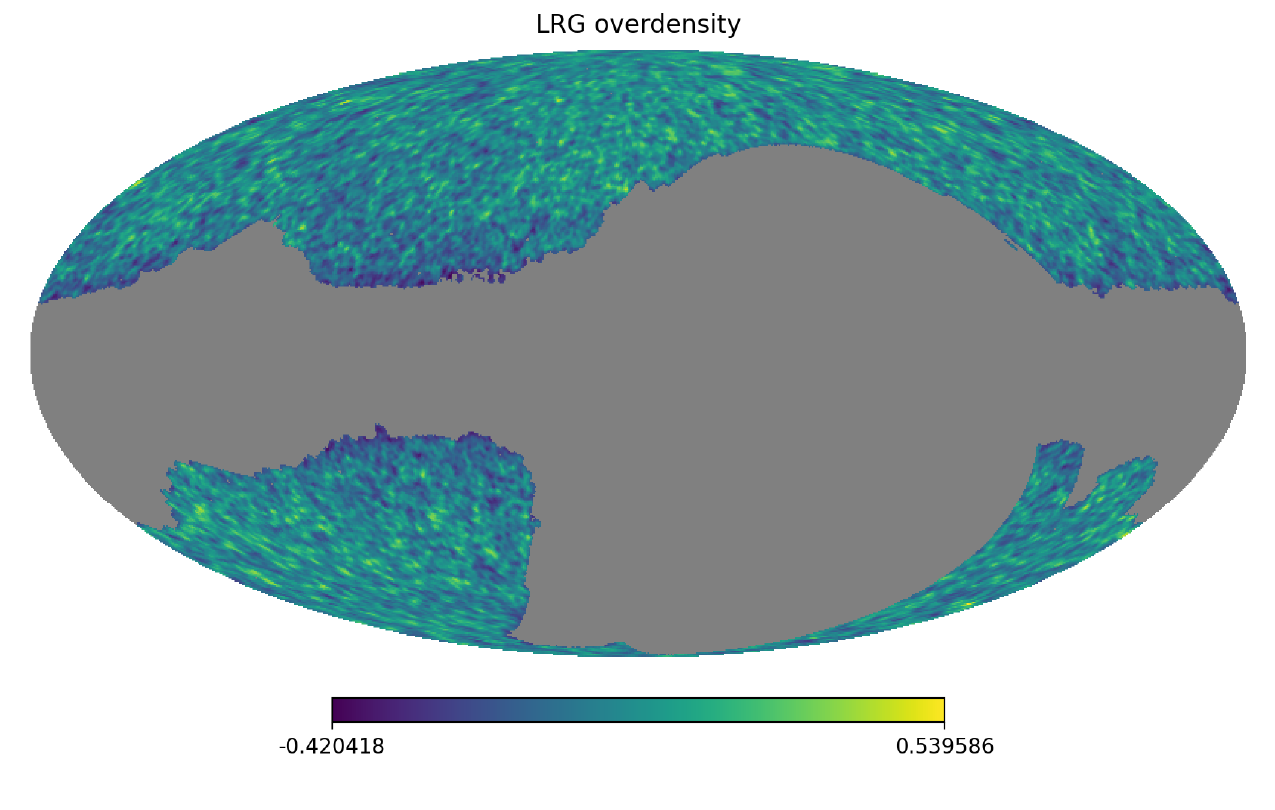}
\includegraphics[width=0.49\textwidth]{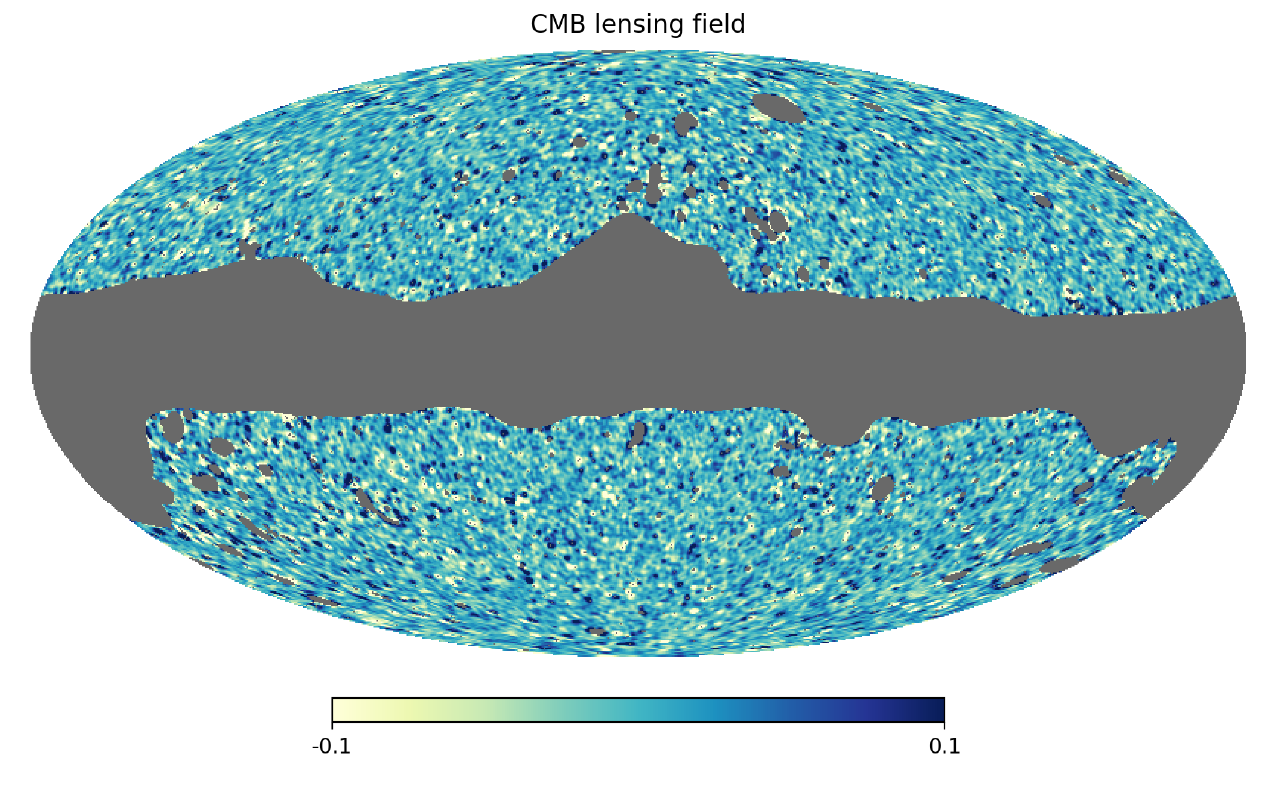}
    \caption{Left panel: LRG overdensity field in galactic coordinates, after applying the $\delta$ < -30º cut. Right panel: CMB lensing field in galactic coordinates, obtained from the $Planck$ PR4 maps.}
    \label{fig:cmb}
\end{figure*}

Our galaxy sample consisted of an LRG catalog obtained from the DESI Imaging Legacy Surveys\footnote{https://www.legacysurvey.org/} DR9 \citep{1804.08657}. These surveys were a combination of three projects: the Dark Energy Camera Legacy Survey using the Blanco 4m telescope in Chile (DECaLS, \citealt{2015AJ....150..150F}), the Beijing-Arizona Sky Survey using the Bok telescope at Kitt Peak (BASS, \citealt{Zou_2017}),  and the Mayall z-band Legacy Survey (MzLS, \citealt{1804.08657}) using the Mayall telescope at Kitt Peak. BASS and MzLS observed the same region in the north galactic cap (NGC), while DECaLS observed in both the NGC and south galactic cap (SGC). The combination of the three projects covered $\sim$19000 deg$^2$ in the sky to select the spectroscopic targets that are currently being observed with DESI. 

We calibrated the redshift distribution, $\dd N/ \dd z$, of the sample using the actual LRG spectra measured with the DESI Survey Validation \citep{DESI2023b.KP1.EDR,DESI2023a.KP1.SV}. In Fig.~\ref{fig:dndz} we show the redshift distribution of the sample as obtained from the LRG spectra. These spectra are not available for declinations lower than $-30^{\circ}$, hence we removed the DEC $< -30^{\circ} $ region from the photometric LRG footprint. \cite{Zhou_2023b} also describe the presence of a photometric zero-point systematic effect at low declinations. The resulting final sample contains around 9 million galaxies covering a $\sim$16000 deg$^2$ area. Then, we applied to the LRG catalog the mask designed by \cite{Zhou_2023} to reduce contamination from effects such as stars and foregrounds. We pixelized the LRG catalog, converting it into a HEALPix \citep{Gorski_2005} galaxy counts map at $N_{\rm side}$ = 256. This map is corrected for the pixel incompleteness effect, which accounts for area losses on scales smaller than a $N_{\rm side} = 256$ HEALpix pixel, such as cutouts around bright stars. Lastly, the galaxy counts map can be easily converted into an overdensity field by normalizing and subtracting the mean density. We show the LRG overdensity field in Fig.~\ref{fig:cmb}.

We note that, since this sample contains photometric redshifts, one could design optimal weights in order to enhance the $f_{\rm NL}$ signal by emphasizing the higher redshift part of the LRG sample. We do not perform this kind of analysis here since we consider it to be beyond the scope of this paper.

\subsection{CMB lensing}
The other ingredient in our analysis was the $Planck$ CMB lensing potential map. We used the $Planck$ PR4 reconstruction of the CMB lensing potential \citep{Carron_2022}, which was obtained from the $Planck$ NPIPE temperature and polarization maps \citep{2020}. In particular, we used the minimum-variance estimate from temperature and polarization, after mean-field subtraction of the lensing convergence. This latest release of CMB lensing maps improves the noise with respect to the previous $Planck$ PR3 maps \citep{1807.06209}; in particular, the large-scale noise is lower and the mean-field is better understood thanks to the larger number of simulations. The maps and the mask are publicly available\footnote{https://github.com/carronj/planck\_PR4\_lensing/releases/tag/Data}. We show the CMB lensing field in Fig~\ref{fig:cmb}.

We note that the CMB lensing map does not have the Monte Carlo multiplicative correction applied in \cite{Carron_2022}. We computed this correction based on simulations as in \cite{Krolewski_2024}, using mode-decoupled pseudo-$C_\ell$, and applied the result as a multiplicative factor to the measured cross-correlation angular power spectra, $C_\ell^{\kappa G}$. This correction cannot be applied in a general way, since it depends on the footprint mask for each tracer involved in the analysis, due to local variations of the normalization. The order of this correction is generally $\lesssim$ 5\%, but it becomes larger ($\sim$10-12\%) for the largest scales, having thus a non-negligible impact on the $f_{\rm NL}$ measurements.

\section{Analysis pipeline}
\label{sec:pipeline}

In this section, we describe the pipeline implemented to analyze the LRG and CMB lensing HEALPix maps. The first step was to apply an imaging systematics mitigation code to the LRG maps. This mitigation treatment operates at the map level and returns a series of systematic weights for each pixel that are applied to the raw LRG maps. Then, we computed the angular power spectrum, $C_\ell$, of the LRGs, their cross-correlation with the $Planck$ lensing, and the covariance matrices. The final step was to perform a Monte Carlo Markov chain (MCMC) based parameter inference to constrain $f_{\rm NL}$.

\begin{figure}
    \centering
\includegraphics[width=.5\textwidth]{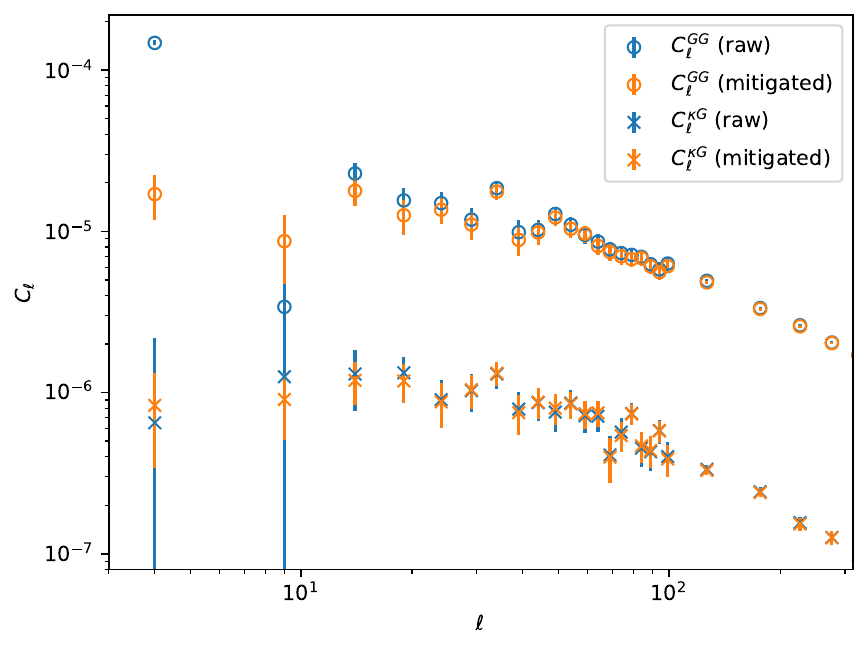}
    \caption{Angular power spectra of the LRG autocorrelation, $C_\ell^{GG}$, and CMB lensing - LRG cross-correlation,  $C_\ell^{\kappa G}$, obtained from the raw (uncorrected) data and after applying the systematics mitigation pipeline to the LRG maps.}
    \label{fig:aps}
\end{figure}
\subsection{Imaging systematics mitigation}
\label{sec:systematics}
Imaging systematics due to effects such as extinction, stellar contamination, or changes in the observational conditions usually generate an excess of power at large scales (low multipoles), where the $f_{\rm NL}$ signal arises. Thus, an efficient imaging systematics treatment is key for measuring an unbiased $f_{\rm NL}$. \cite{2020MNRAS.495.1613R} present a neural network approach for systematics mitigation  as a way to model the relation between the observed galaxy density fields and the imaging systematics templates. This pipeline is implemented in the \texttt{SYSnet} code, which is publicly available\footnote{https://github.com/mehdirezaie/SYSNet}. In \cite{rezaie2023local}, a detailed study of the performance of \texttt{SYSnet} was done using this LRG sample, with the aim of measuring $f_{\rm NL}$ from the $C_\ell^{GG}$ autospectra. A different number of approaches were explored, given that \texttt{SYSnet} can return different results depending on the selection of features (imaging systematics template maps) used to perform the regression. 

In this work, we used a mitigation recipe optimized for measurements from the cross-correlation between the DESI LRG and $Planck$ lensing, $C_\ell^{\kappa G}$. At first order, the cross-correlation itself as a technique would be enough to remove the effects of the systematics if we assumed there were no correlated systematics between the CMB lensing and LRG maps. However, there could still be potential correlated systematics due to galactic foregrounds that affect both probes. Furthermore, the noise in $C_\ell^{GG}$ contributes to the $C_\ell^{\kappa G}$ covariance matrix: systematics affecting the $C_\ell^{GG}$ power spectrum would also lead to an increase in the $C_\ell^{\kappa G}$ covariance, and as a result, larger uncertainties on $f_{\rm NL}$. At the same time, a very aggressive mitigation recipe could remove real clustering signal and overfit the $C_\ell^{\kappa G}$ power spectrum, biasing the $f_{\rm NL}$ measurements toward lower values. For this reason, in order to find a compromise between removing the excess of power spectrum due to systematics for $C_\ell^{GG}$ and avoiding a strong overfit of $C_\ell^{\kappa G}$, we selected the ``nonlinear three maps'' recipe from \cite{rezaie2023local}. This choice is the most conservative one among the recipes applied in \cite{rezaie2023local} with the \texttt{SYSnet} code, and relies on selecting three features or systematics templates for the regression: extinction, galactic depth in the $z$-band, and point spread function (PSF) size in the $r$-band. We refer the reader to \cite{rezaie2023local} for more details on the different possibilities of feature selection. In Sect.~\ref{sec:mocks} we also show that this prescription provides unbiased measurements of the angular power spectrum on contaminated mocks.

\subsection{Angular power spectra and covariance matrix}
\label{sec:aps}

\begin{figure}
    \centering
\includegraphics[width=.5\textwidth]{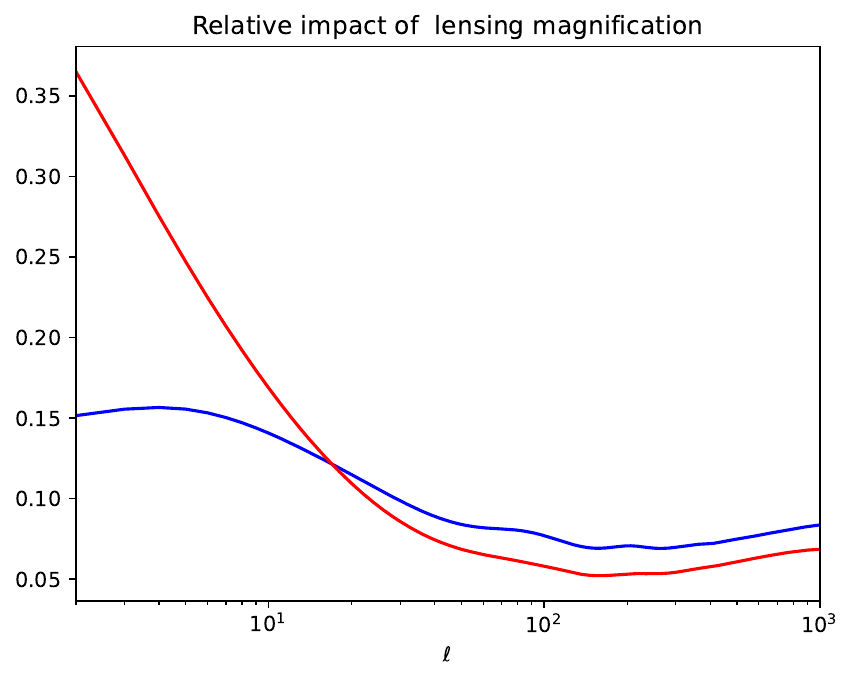}
    \caption{Relative impact of the lensing magnification contributions for the $C_\ell^{GG}$  (blue line) and $C_\ell^{\kappa G}$ (red line) theoretical angular power spectra obtained with \texttt{CAMB}.}
    \label{fig:sz}
\end{figure}

In order to estimate the angular power spectra of the LRG autocorrelation and their cross-correlation with the $Planck$ lensing, we used the pseudo-$C_\ell$ approach implemented in the publicly available \texttt{NaMaster} code by \citet{1809.09603}. The pseudo-$C_\ell$ of a pair of fields can be defined as
\begin{equation}
    \Tilde{C}_\ell^{XY} = \frac{1}{2 \ell + 1} \sum_m X_{\ell m} Y_{\ell m}^{*},
\end{equation}
where $X$ and $Y$ are the observed fields. Then, the difference between the true and measured $C_\ell$ due to the effects of the mask is accounted for through the mode-coupling matrix $M_{\ell \ell'}$ as
\begin{equation}
    \langle \Tilde{C}_\ell \rangle = \sum_{\ell'} M_{\ell \ell'} C_{\ell'} \,.
\end{equation}
 In our case, we directly applied a completeness mask to the observed LRG density, so we just deconvolved the binary footprint mask. In practice, the inversion of the $M_{\ell \ell'}$ matrix is done using the MASTER algorithm \citep{Hivon_2002}, which requires a discrete binning of the angular power spectrum. We used the implementation in the \texttt{compute\_full\_master} function from  the \texttt{NaMaster} code to calculate $C_\ell^{\kappa G}$ and $C_\ell^{GG}$. We binned the theory curves using the same bandpower window functions.

As scale cuts, for $ C_\ell^{\kappa G}$we adopted $\ell_{\rm min}$ = 2 and $\ell_{\rm max}$ = 300, and we binned the power spectra using $\Delta \ell$ = 5 for $\ell < 100$ and $\Delta \ell$ = 50 for $\ell > 100$. The reason for using this scheme was to obtain a good sampling of the angular power spectra at large scales where the $f_{\rm NL}$ signal arises and $\ell_{\rm max}$ = 300 is safe enough to minimize the limitations of modeling the nonlinear scales, which could potentially affect the measurement of the linear bias. We checked, with mock fields, that the results are stable with both $N_{\rm side}$ and $\ell_{\rm max}$ (see Sect.~\ref{sec:mocks} for more details on the mocks). For $C_\ell^{GG}$, we implemented the same multipole binning scheme, but we dropped the first multipole bin of the analysis and set $\ell_{\rm min}$ = 7. The motivation for this choice was to use a scale cut for $C_\ell^{GG}$ that limits the impact of remaining systematics according to the mitigation pipeline tests on mocks (see Sect.~\ref{sec:mocks} for more details).
\begin{figure}
    \centering
\includegraphics[width=.5\textwidth]{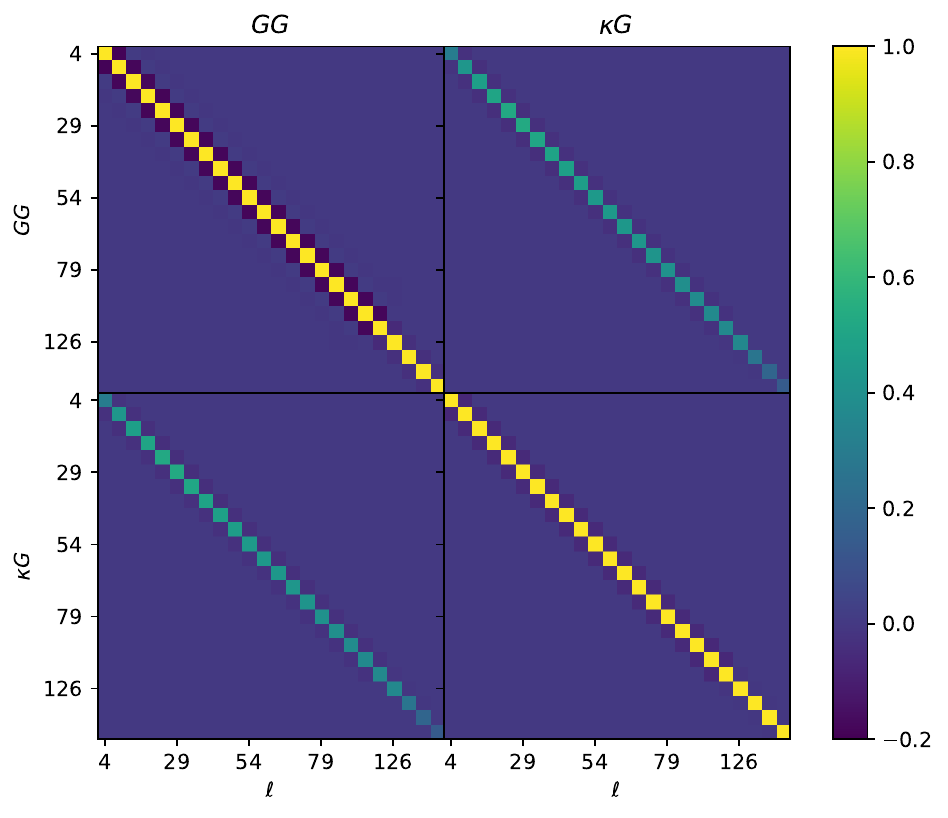}
    \caption{Correlation matrix as obtained from the joint covariance matrix for $C_\ell^{GG}$ and $C_\ell^{\kappa G}$ computed with the \texttt{NaMaster} code.}
    \label{fig:covmat}
\end{figure}
We show in Fig.~\ref{fig:aps} the measured power spectra of the LRG autocorrelation, $C_\ell^{GG}$, and its cross-correlation with the $Planck$ lensing, $C_\ell^{\kappa G}$, obtained using uncorrected (raw) LRG maps and LRG maps mitigated with the neural network pipeline.
For the covariance matrices, we used the analytic \texttt{gaussian\_covariance} function in \texttt{NaMaster}  \citep{GarciaGarcia19} to compute the full Gaussian covariance for a masked field. We accounted for the potential extra power in $C_\ell^{GG}$ due to systematics by smoothing the measured angular power spectra from the data and using this as input for the covariance matrix computation. As a test, we also computed the covariance using mock fields, obtaining compatible results. More details of the mock fields used can be found in Sect.~\ref{sec:mocks}. We show in Fig.~\ref{fig:covmat} the computed joint correlation matrix for $C_\ell^{GG}$ and $C_\ell^{\kappa G}$.

\subsection{Likelihood and parameter inference}
\label{sec:likelihood}
We defined the likelihood, ${\cal L}$, as
\begin{equation}
   -2 \log {\cal L} \equiv \chi^2 = \sum_{\ell,\ell'} \left(C_\ell^{{\rm obs}} - \Tilde{C}_\ell (\theta) \right){\rm Cov}^{-1}_{\ell \ell'}\left(C_{\ell'}^{{\rm obs}} - \Tilde{C}_{\ell'} (\theta) \right),
\end{equation}
where $C_\ell^{{\rm obs}}$ are the elements of the data vector, $\Tilde{C}_\ell (\theta)$ is the theoretical model of the angular power spectrum for a given set of parameters, $\theta$, and ${\rm Cov}^{-1}$ is the inverse of the covariance matrix.

\begin{figure*}
    \centering
\includegraphics[width=0.49\textwidth]{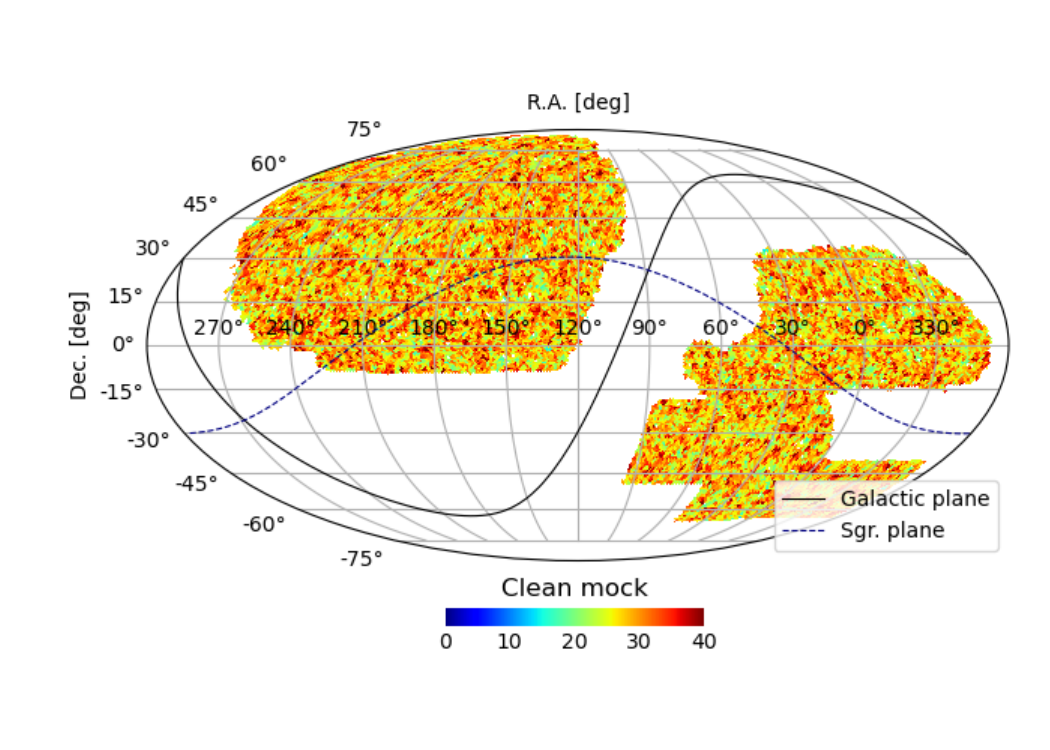}
\includegraphics[width=0.49\textwidth]{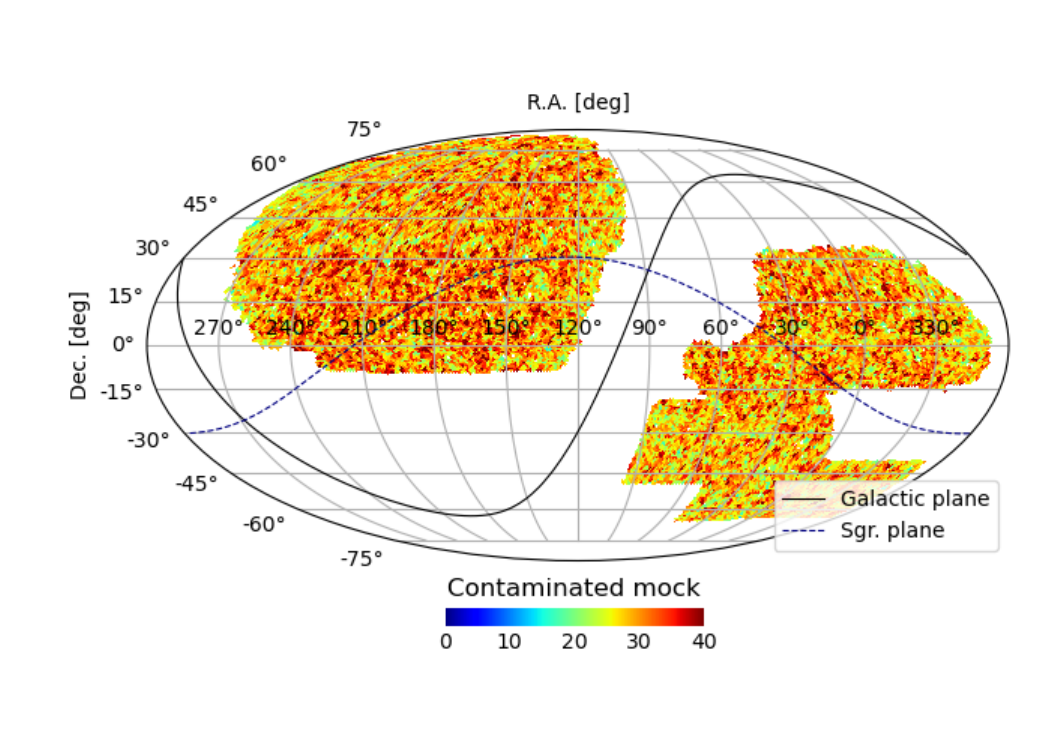}
    \caption{Single Gaussian realization of an LRG map before adding contamination (left panel) and after applying \texttt{regressis} to contaminate the mock (right panel).}
    \label{fig:contmap}
\end{figure*}

\begin{figure*}[ht!]
    \centering
\includegraphics[width=0.49\textwidth]{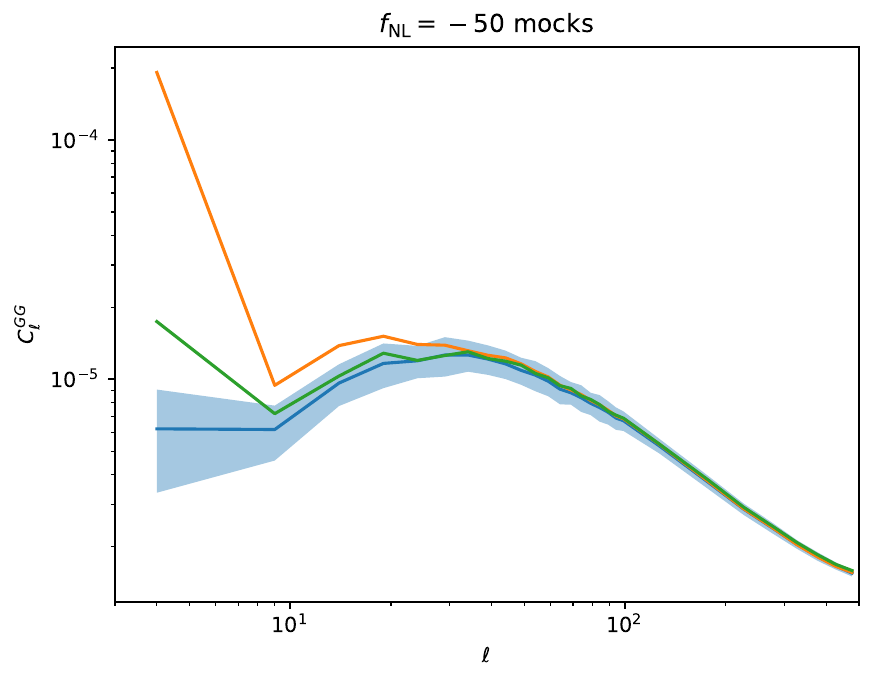}
\includegraphics[width=0.49\textwidth]{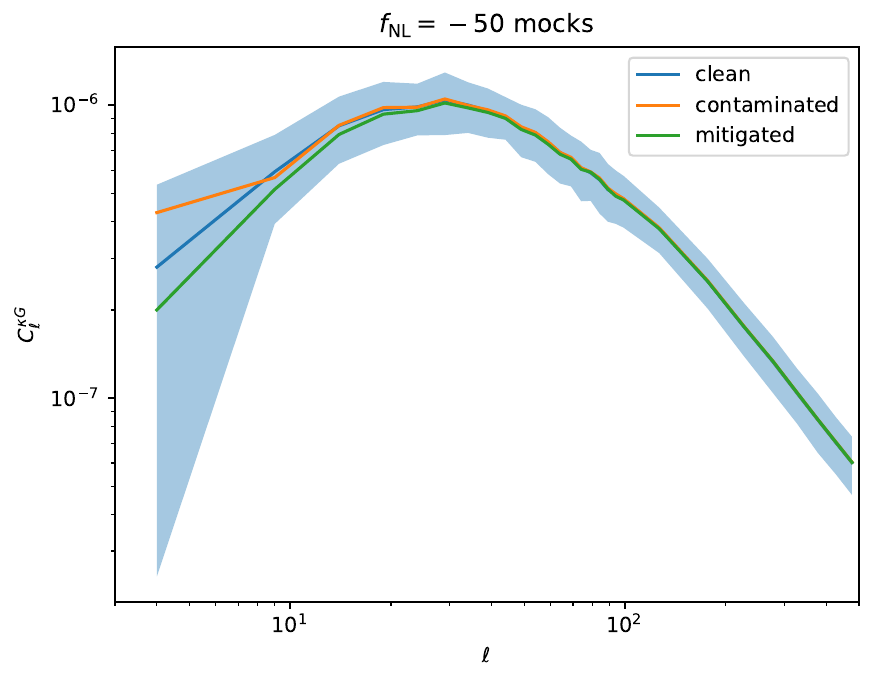}
\includegraphics[width=0.49\textwidth]{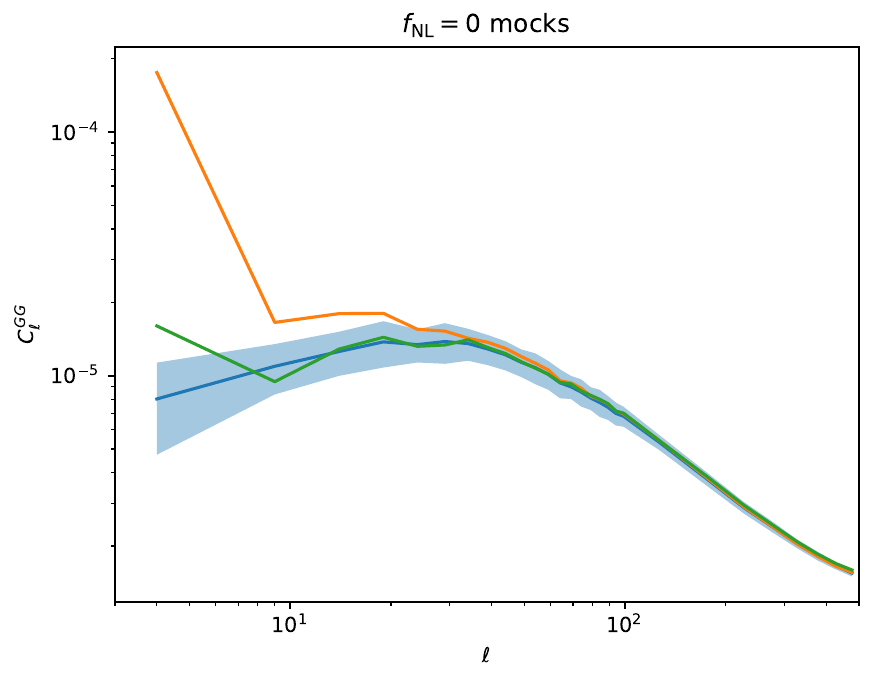}
\includegraphics[width=0.49\textwidth]{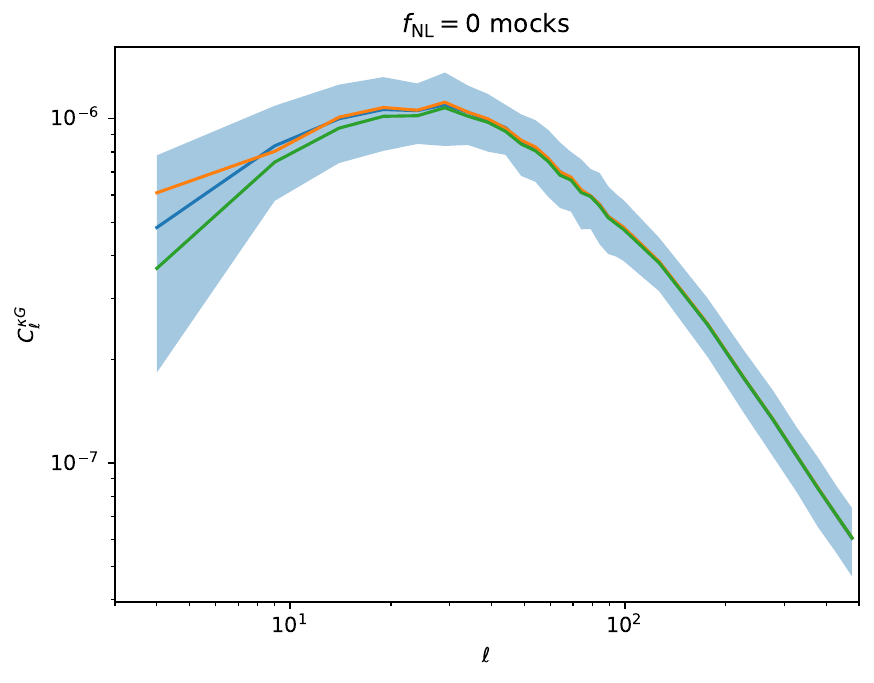}
\includegraphics[width=0.49\textwidth]{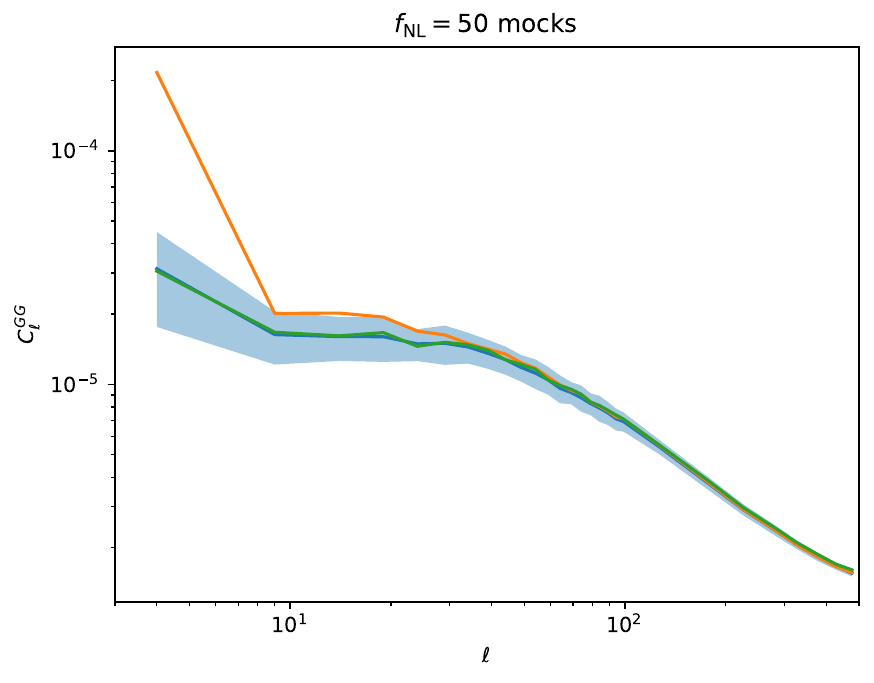}
\includegraphics[width=0.49\textwidth]{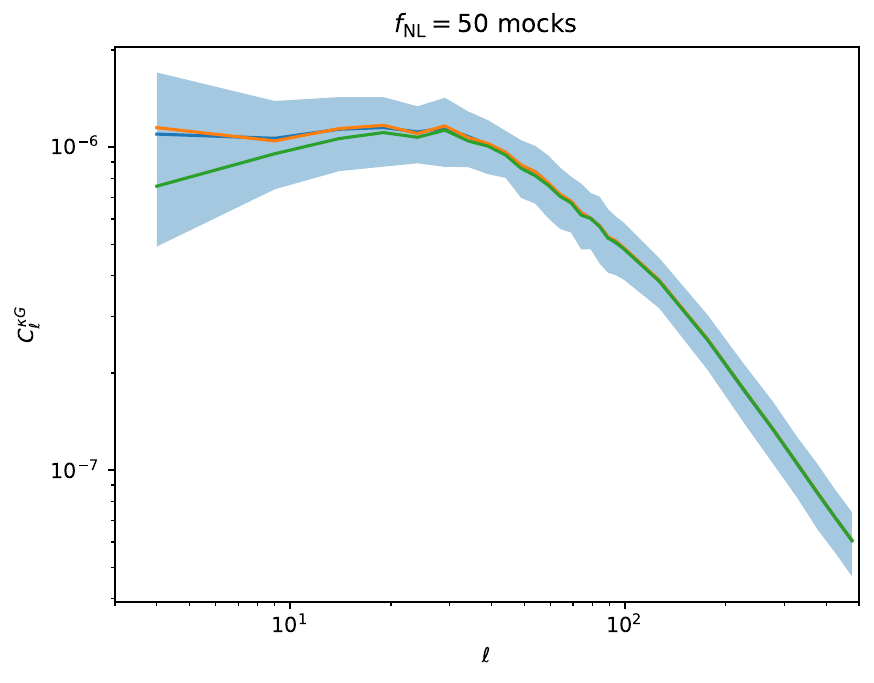}
    \caption{Mean angular power spectra of the full sample LRG autocorrelation (left panels) and LRG-CMB lensing cross-correlation (right panels), computed from the 100 mock realizations used to test our pipeline for various values of $f_{\rm NL}$ (-50 top panels, 0 middle panels, and 50 bottom panels). The blue lines show the true angular power spectra before adding contamination, the shaded blue area corresponds to the dispersion on the true spectra, the orange lines correspond to the angular power spectra of the contaminated mocks, and the green lines to the contaminated mocks after imaging systematics mitigation.}
    \label{fig:apsmocks}
\end{figure*}

Our theoretical model is based on the Code for Anisotropies in the Microwave Background, \texttt{CAMB} \citep{2011ascl.soft02026L}. This code is publicly available\footnote{https://camb.info/} and allows the computation of the angular power spectra of the CMB fields and also galaxy fields, given a window function determined by the galaxy bias, $b_g$, and the redshift distribution, $\dd N/\dd z$. We slightly modified the code to include the scale dependence of the galaxy bias induced by $f_{\rm NL}$. The $\dd N/\dd z$ was directly measured using the DESI LRG spectra from the Survey Validation \citep{DESI2023b.KP1.EDR,DESI2023a.KP1.SV}. The rest of the fiducial cosmological parameters in \texttt{CAMB} are fixed to the $Planck$ 2018 best-fit estimates \citep{1807.06209}, except for $\sigma_8$, which we set to a fiducial value $\sigma_8 = 0.77$ to be in agreement with the measurements from the cross-correlation between the Atacama Cosmology Telescope (ACT) lensing and this LRG sample \citep{sailer2024cosmologicalconstraintscrosscorrelationdesi,kim2024atacamacosmologytelescopedr6}. 

In the computation of the theoretical angular power spectra with the \texttt{CAMB} code, we included the lensing magnification contribution to the galaxy number counts. This effect accounts for the fact that the light from distant galaxies is affected by structures along the line of sight, resulting in an increased flux. The lensing magnification contribution to the angular power spectra could be included in \texttt{CAMB} once we specified the magnification bias, $s$, as an input parameter. This parameter depends on the galaxy tracer, and for this LRG sample we fixed the magnification bias value to $s = 0.999$, as determined in \cite{2010.04698}. More recent measurements using this sample split into four redshift bins \citep{2111.09898,Zhou_2023b} find compatible results within the uncertainty for the various redshift bins; hence, we considered as safe enough the assumption of a $z$-independent value for $s$. In Fig.~\ref{fig:sz} we show the relative importance of the lensing magnification contribution to the $C_\ell^{GG}$ and $C_\ell^{\kappa G}$ theoretical angular power spectra, which can be expressed as $C_\ell^{XX, \rm density + lensing} / C_\ell^{XX,\rm density} - 1$. This was calculated with \texttt{CAMB} by adding or neglecting the contribution to the galaxy number counts described in Eq.~\ref{eq:deltalens}. The impact at the lowest multipoles can reach up to $\sim$ 15\% for the LRG autocorrelation and $\sim$ 35\% for the LRG - CMB lensing cross-correlation; hence, this effect is not negligible in our theoretical model for this tracer.

To compute the constraints on the cosmological parameters, we implemented our likelihood using the MCMC sampler \texttt{emcee}\footnote{https://emcee.readthedocs.io/} \citep{Foreman_Mackey_2013}. Our analysis includes $f_{\rm NL}$ and the galaxy bias at $z=0$, $b_0$, as the two main cosmological parameters of interest to constrain. We assumed a fiducial redshift evolution of the galaxy bias following
\begin{equation}
\label{eq:bz}
b_g = b_0 \times D(z)^{-1},
\end{equation}
where $D(z)$ is the growth factor normalized to be 1 at $z=0$. This choice was motivated by the analysis in \cite{2001.06018}, where a bias evolution compatible with $b(z) \simeq 1.5/D(z)$ was found for the DESI LRG targets.
For $C_\ell^{GG}$, we also considered the shot noise, $N_{\rm shot}$, as a nuisance parameter. We did not impose any prior knowledge on the $b_0$ and $N_{\rm shot}$ parameters, but restricted their sampling to positive values to avoid nonphysical results.

\section{Validation with mocks}
\label{sec:mocks}

\label{sec:mocks}
Before applying the analysis pipeline described in the previous section to the real LRG and CMB lensing data, we tested it with mock Gaussian fields that simulated the DESI Legacy Survey LRG sample and the $Planck$ lensing observations. For the LRG, we used a sample of 100 Gaussian fields for $f_{\rm NL}$ = 0, 50, and -50. For the CMB lensing, we used a set of 100 correlated maps. These correlated fields were generated using the \texttt{healpy}\footnote{https://github.com/healpy} Python package.

We first used \texttt{CAMB} to compute the theoretical angular power spectra $C_\ell^{GG}$, $C_\ell^{\kappa G}$, and $C_\ell^{\kappa \kappa}$, given the redshift window function d$N$/d$z$ obtained from the spectroscopic LRG redshift distribution and the different values of $f_{\rm NL}$. The galaxy bias parameter at $z=0$ was set to a fiducial value of $b_0 = 1.5$, based on \cite{2001.06018}. The fiducial cosmological parameters were set according to the $Planck$ 2018 measurements \citep{1807.06209}.

 The theoretical angular power spectra were used as input for the \texttt{healpy.synfast} function. Following the recipe in \cite{Giannantonio_2008,1404.1933} for simulating correlated fields, we first built a CMB lensing map with a given seed and power spectrum, $C_\ell^{\kappa \kappa}$. Then, we generated another map with the same seed and power spectrum, $(C_\ell^{\kappa G})^2/C_\ell^{\kappa \kappa}$, and we added to this last map another component with a different seed and power spectrum, $C_\ell^{GG} - (C_\ell^{\kappa G})^2/C_\ell^{\kappa \kappa}$. These maps will have amplitudes given by
\begin{equation}
\begin{aligned}
& a_{\ell m}^{\kappa \kappa} = \xi_1 (C_\ell^{\kappa \kappa})^{1/2}, \\
& a_{\ell m}^{GG} = \xi_1 C_\ell^{\kappa G}/(C_\ell^{\kappa \kappa})^{1/2} + \xi_2\left(C_\ell^{GG} - (C_\ell^{\kappa G})^2/C_\ell^{\kappa \kappa}\right)^{1/2},   
\end{aligned}     
\end{equation}
where $\xi_1$ and $\xi_2$ are random amplitudes, that is, complex numbers with zero mean and unity variance. It is easy to show that
\begin{equation}
\begin{aligned}
  &  \langle a_{\ell m}^{\kappa \kappa} a_{\ell m}^{\kappa \kappa *} \rangle = C_\ell^{\kappa \kappa}, \\
  &  \langle a_{\ell m}^{\kappa \kappa} a_{\ell m}^{GG *} \rangle = C_\ell^{\kappa G}, \\
  &  \langle a_{\ell m}^{GG} a_{\ell m}^{GG *} \rangle = C_\ell^{GG}.
\end{aligned}
\end{equation}
The output products are a subsample of 100 mock CMB lensing fields and 100 correlated LRG overdensity fields, with $N_{\rm side} = 256$ for each value of $f_{\rm NL}$. Lastly, we applied to the mocks the same masks we used for the LRG and CMB lensing data.

In order to test the performance of the systematics mitigation pipeline, we contaminated the LRG maps. For this purpose, after generating the mock LRG and CMB lensing fields, we used the \texttt{regressis}\footnote{https://github.com/echaussidon/regressis} code \citep{10.1093/mnras/stab3252}. From the LRG density fields, we simulated LRG discrete number count maps using Poisson sampling, based on the expected LRG density for each pixel. This procedure was safe enough due to the low map resolution, which made it unlikely we would get pixels with predicted density lower than 0 for the Gaussian fields. We also checked the angular power spectra were in agreement with the input after Poisson sampling. We then generated ``high density'' simulated LRG maps by duplicating the number of objects per pixel. Then, we created a mock LRG catalog by assigning random coordinates within a given pixel to each galaxy. Finally, the ``high density'' catalog was used as input for \texttt{regressis}. Based on the systematics weights,  which were estimated from running SYSnet on the real data, the code created a contaminated mock catalog by removing galaxies and matching the final catalog to the expected LRG density. The result was converted back into a contaminated map.
    
We note that we did not add any contamination to the CMB lensing mock fields. This was equivalent to assuming there is no correlation in systematics between the two probes, and the main purpose of this test was to check whether the systematics mitigation removed real cross-correlation signal in the $C_\ell^{\kappa G}$ power spectra.

We show in Fig.~\ref{fig:contmap} an example of a clean mock map and the same mock map contaminated after applying \texttt{regressis}. These contaminated maps were used as input for \texttt{SYSNet}, the neural network code for systematics mitigation described in Sect. ~\ref{sec:systematics}. For each realization, we applied \texttt{SYSNet} separately in three different regions of the sky: one corresponding to the BASS and MzLS footprints and two corresponding to DECaLS in the NGC and SGC (see Fig. 2 of \citealt{rezaie2023local}). After recombining the output in a single map, we computed the angular power spectra of the lensing - LRG cross-correlation, $C_\ell^{\kappa G}$, and the LRG autocorrelation, $C_\ell^{GG}$, for the mocks without the contamination (clean power spectrum), then we added the contamination and finally applied the systematics weights obtained with \texttt{SYSNet}.

The comparison of angular power spectra is shown in Fig.~\ref{fig:apsmocks}. For 
        $C_\ell^{\kappa G}$, we find no significant differences between the contaminated and true power spectra. While the mitigated spectra are systematically lower than the truth, this difference is smaller than the dispersion of the clean mock power spectra (i.e.,\ the $C_{\ell}^{\kappa g}$ uncertainty). In relative terms, at the lowest multipoles where the $f_{\rm NL}$ signal arises, we find differences between the mitigated and true spectra of up to $\sim$15\%, $\sim$25\%, and $\sim$30-35\% for the $f_{\rm NL} = -50$, 0, and 50 mocks, respectively.
        Nonetheless, constraining $f_{\rm NL}$ from $C_\ell^{\kappa G}$ using uncorrected maps would lead to a much larger $C_{\ell}^{\kappa g}$
        covariance from 
        the large amount of extra power in $C_\ell^{GG}$, significantly increasing the $f_{\rm NL}$ uncertainty. For this reason, we used the mitigated maps as the baseline in our cross-correlation pipeline.

For $C_\ell^{GG}$, the systematics mitigation presents a more complicated picture: we find that for the $f_{\rm NL} = 50$ mocks the mitigated spectra are compatible with the dispersion of the true spectra, but the performance of the mitigation pipeline depends on $f_{\rm NL}$: for $f_{\rm NL} = -50$ and $f_{\rm NL} = 0$, the first multipole bin of the mitigated spectra is clearly
still too high, lying outside the 1$\sigma$ range of the clean mocks. Thus, for $C_\ell^{GG}$ we imposed a scale cut of the first five multipoles contained in this bin and adopted $\ell_{\rm min} = 7$ for the parameter inference with the mocks and real data. Having a larger $\ell_{\rm min}$ for $C_\ell^{GG}$ than for $C_\ell^{\kappa G}$ also allowed us to put a joint constraint on $f_{\rm NL}$, while having better control of the systematics. 

We then applied the MCMC parameter inference pipeline on the mean of the clean and mitigated angular power spectra of the 100 mock realizations for each value of $f_{\rm NL}$. We analyzed $C_\ell^{\kappa G}$ and $C_\ell^{GG}$ separately and jointly in order to understand the impact of each observable on the constraints, as well as the effects of possible remaining systematics. We list the median likelihood values with 68\% confidence intervals in Tab.~\ref{tab:cleanmocks} for the mocks without systematics (clean) and in Tab.~\ref{tab:mocks} for the mocks with systematics (mitigated). For the clean mocks, we find almost no intrinsic bias on the recovered $f_{\rm NL}$ values, always having a $\lesssim 0.3 \sigma$ agreement with the input. For the constraints from $C_\ell^{\kappa G}$ on mocks after contamination and systematics mitigation, the input $f_{\rm NL}$ is always within the 1$\sigma$ error bars of the measurements on mocks, and the bias on the measured $f_{\rm NL}$ is also lower than 0.5$\sigma$ for the $f_{\rm NL} = 0$ and $f_{\rm NL} = -50$ mocks. The larger disagreement between the measured and true values for the $f_{\rm NL} = 50$ mocks can be interpreted as a consequence of an overfitted power spectrum after applying the systematics mitigation pipeline (with the bottom right panel of Fig.~\ref{fig:apsmocks}
showing the largest offset between mitigated and clean in the lowest $\ell$ bin, in units of the standard deviation). For $C_\ell^{GG}$, we also find a level of agreement between the measured and true $f_{\rm NL}$ values within 1$\sigma$; however, in this case, the highest bias on the  measurement happens for the $f_{\rm NL} = -50$ mocks. For the joint $C_\ell^{\kappa G}+C_\ell^{G G}$ tests, using $\ell_{\rm min}$ = 7 for $C_\ell^{G G}$, we find a good agreement for the zero and positive $f_{\rm NL}$ cases, and a $\lesssim$1$\sigma$ bias on the $f_{\rm NL} = -50$ mocks. Additionally, due to the systematics uncertainties on $C_\ell^{G G}$, we measured the constraints from mocks using a joint approach in which we kept the $C_\ell^{G G}$ information for $\ell > 32$ only, in such a way that it constrained the galaxy bias but did not affect the scales where the $f_{\rm NL}$ signal arose. 

The results of the tests on mocks with systematics reported in Tab.~\ref{tab:mocks}
are driven by the impact of the systematics mitigation on the angular power spectra shown in Fig.~\ref{fig:apsmocks}: for the negative $f_{\rm NL}$ mocks, there is a quite limited ($\sim$15\%) overfit on $C_\ell^{\kappa G}$ and some remaining power for $C_\ell^{GG}$ , while for positive $f_{\rm NL}$ mocks there is a stronger overfit ($\gtrsim$ 30\%) on $C_\ell^{\kappa G}$ and the true power spectrum is almost perfectly recovered for $C_\ell^{G G}$. In consequence, for the negative $f_{\rm NL}$ mocks, the constraints from $C_\ell^{\kappa G}$ present almost no bias with respect to the true $f_{\rm NL}$, while the constraints from $C_\ell^{G G}$ are biased by $\lesssim 1 \sigma$ toward higher $f_{\rm NL}$ values, likely due to remaining systematics. The opposite behavior is found for the positive $f_{\rm NL}$ mocks: the constraints from $C_\ell^{\kappa G}$ are biased by $\lesssim 1 \sigma$, while the constraints from $C_\ell^{G G}$ present a better accuracy. For the $f_{\rm NL}$ = 0 mocks, we find almost no bias for $C_\ell^{G G}$ and a 0.5$\sigma$ bias for $C_\ell^{\kappa G}$, also driven by the overfit of systematics mitigation. In the case of the joint constraints, when adopting the baseline $\ell_{\rm min} = 7$ for $C_\ell^{G G}$, the constraints behave in a similar way to those from the autocorrelation, performing accurately for zero and positive $f_{\rm NL}$. Our results show that, even if there could be possible biases on $f_{\rm NL}$ due to remaining or overcorrected systematics, the input $f_{\rm NL}$ is within the 1$\sigma$ uncertainties of the measurement for every single case.

\begin{table}
    \centering
    \begin{tabular}{|c|c|c|c|c|}
    \hline
         & $C_\ell^{\kappa G}$  & $C_\ell^{GG}$ & $C_\ell^{\kappa G}$+$C_\ell^{GG}$ & $C_\ell^{\kappa G}$+$C_\ell^{GG}$ \\ \hline
        Scale cut & $\ell_{\rm min} = 2$ & $\ell_{\rm min} = 7$ & $\ell_{\rm min} = 2, 7$ & $\ell_{\rm min} = 2, 32$  \\ \hline
         $f_{\rm NL}^{\rm true}$ = -50 & -50 $\pm$ 30 & -43$_{-23}^{+22}$ & -45 $\pm$ 21 & -39 $\pm$ 27\\ \hline
        $f_{\rm NL}^{\rm true}$ = 0 & 5$_{-35}^{+31}$ & 3$^{+20}_{-21}$ & 5 $\pm$ 19  & 9$_{-27}^{+30}$ \\ \hline
         $f_{\rm NL}^{\rm true}$ = 50 & 49$_{-34}^{+35}$  & 45$^{+21}_{-20}$ & 49$^{+16}_{-17}$ & 58$^{+31}_{-32}$ \\ \hline     
    \end{tabular}
    \caption{Median likelihood $f_{\rm NL}$ values with 68\% confidence intervals obtained from the application of our analysis pipeline to clean Gaussian mocks with $f_{\rm NL}$ = -50, 0, and 50. }
    \label{tab:cleanmocks}
\end{table}

\begin{table}
    \centering
    \begin{tabular}{|c|c|c|c|c|}
    \hline
         & $C_\ell^{\kappa G}$  & $C_\ell^{GG}$ & $C_\ell^{\kappa G}$+$C_\ell^{GG}$ & $C_\ell^{\kappa G}$+$C_\ell^{GG}$ \\ \hline
        Scale cut & $\ell_{\rm min} = 2$ & $\ell_{\rm min} = 7$ & $\ell_{\rm min} = 2, 7$ & $\ell_{\rm min} = 2, 32$  \\ \hline
         $f_{\rm NL}^{\rm true}$ = -50 & -48$^{+34}_{-31}$ & -34$_{-27}^{+22}$ & -30$_{-21}^{+19}$ & -30 $\pm$ 27\\ \hline
         $f_{\rm NL}^{\rm true}$ = 0 & -17$^{+37}_{-35}$ & -4$^{+24}_{-23}$ & -1$^{+17}_{-18}$ & -3$_{-28}^{+30}$\\ \hline
         $f_{\rm NL}^{\rm true}$ = 50 & 21 $\pm$ 32  & 48$^{+17}_{-19}$ & 46$^{+15}_{-16}$  & 37 $\pm$ 28  \\ \hline     
    \end{tabular}
    \caption{Median likelihood $f_{\rm NL}$ values with 68\% confidence intervals obtained from the application of our analysis pipeline to contaminated and mitigated Gaussian mocks with $f_{\rm NL}$ = -50, 0, and 50.}
    \label{tab:mocks}
    \tablefoot{The constraints from $C_\ell^{GG}$ used a scale cut at $\ell_{\rm min} = 7$ to have a better control of systematics, while the constraints from $C_\ell^{\kappa G}$ were obtained using all the multipoles up to $\ell_{\rm max} = 300$. The joint constraints also include a case in which the scale cut for $C_\ell^{GG}$ was set to $\ell_{\rm min} = 32$  in order to constrain only the galaxy bias but not $f_{\rm NL}$.}

\end{table}

\section{Results}
\label{sec:results}
\begin{figure}
    \centering
\includegraphics[width=0.5\textwidth]{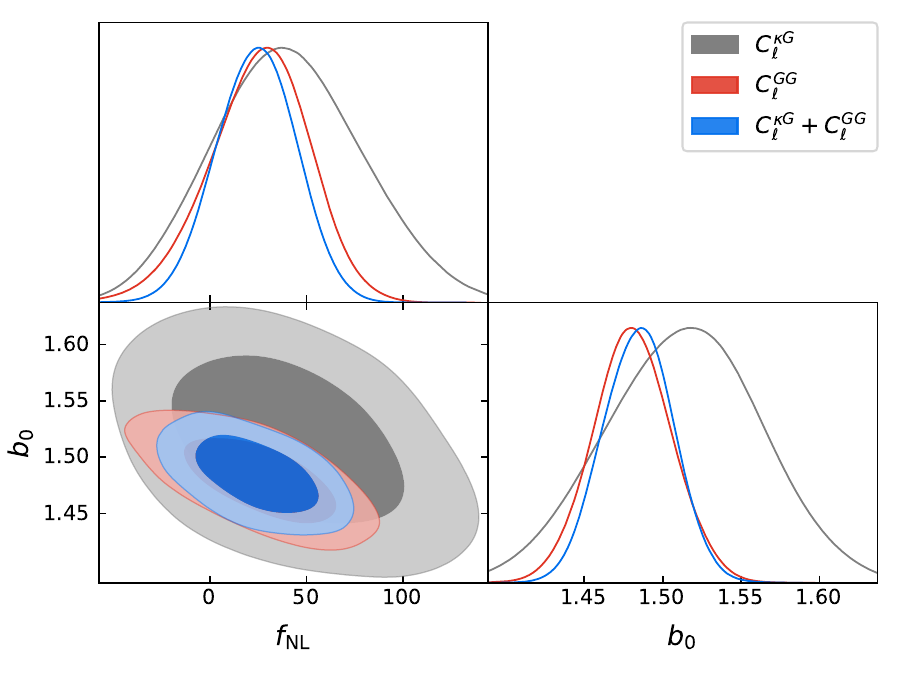}
    \caption{1$\sigma$ and 2$\sigma$ confidence ellipses for the joint posterior distribution of the $f_{\rm NL}$ and $b_0$ parameters obtained from the $C_\ell^{\kappa G}$ cross-correlation (grey contours), the $C_\ell^{G G}$ autocorrelation (red contours), and both observables jointly (blue contours).}
    \label{fig:results}
\end{figure}

We list in Tab.~\ref{tab:results} the median likelihood values with 68\% confidence intervals for $f_{\rm NL}$ and $b_0$, obtained from the MCMC analysis applied to $C_\ell^{\kappa G}$ and $C_\ell^{GG}$ separately and jointly. We obtain an uncertainty $\sigma(f_{\rm NL}) \lesssim$ 40 when using the $C_\ell^{\kappa G}$ cross-correlation only. Using a scale cut at $\ell_{\rm min} = 7$, this uncertainty is reduced to $\sigma(f_{\rm NL}) \sim$ 25 from $C_\ell^{G G}$ only and to $\sigma(f_{\rm NL}) \sim$ 20 from the combination of both observables. The median likelihood values of $f_{\rm NL}$ suggest a preference of this LRG sample for positive values of $f_{\rm NL}$, which could be interpreted as a consequence of remaining systematics. However, taking into account the error bars, our results are consistent with a $\Lambda$CDM universe with Gaussian initial conditions at a $\sim 1\sigma$ level for  $C_\ell^{\kappa G}$ and $\gtrsim 1 \sigma$ when combining this information with the LRG autospectra, $C_\ell^{G G}$. For the galaxy bias parameter, $b_0$, we obtain a compatible result with previous measurements using this LRG sample: for example, \cite{2001.06018} measured $b_0 \sim 1.5$ assuming a redshift evolution proportional to $D(z)^{-1}$. Fig.~\ref{fig:results} shows the 68\% and 95\% confidence ellipses for the $f_{\rm NL}$ - $b_0$ plane obtained from the two observables and their combination.

\begin{figure}
    \centering
\includegraphics[width=0.5\textwidth]{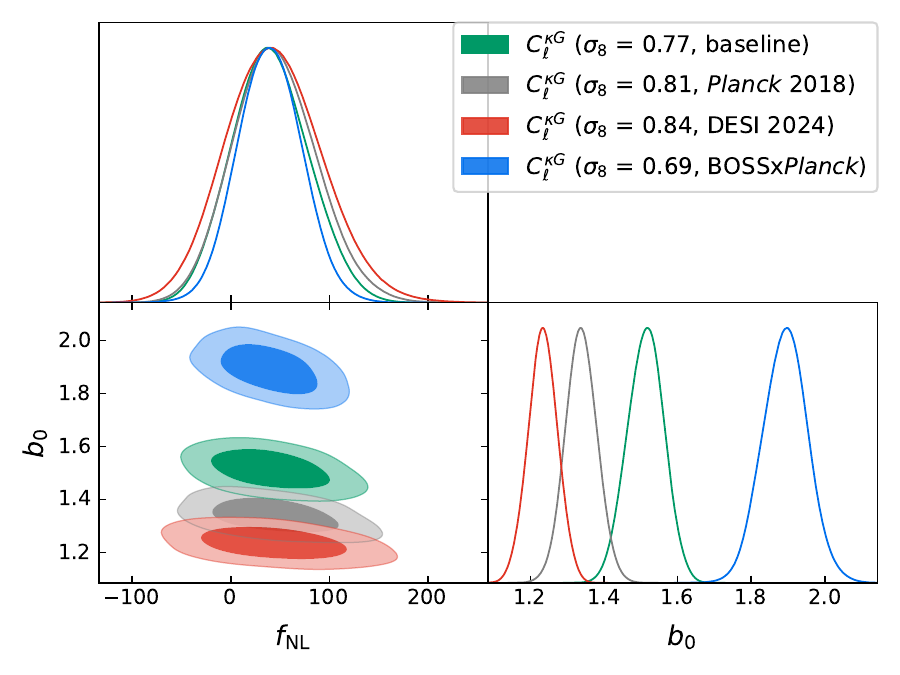}
    \caption{1$\sigma$ and 2$\sigma$ confidence ellipses for the joint posterior distribution of the $f_{\rm NL}$ and $b_0$ parameters obtained from the $C_\ell^{\kappa G}$ cross-correlation. The green contours assume the baseline $\sigma_8$ value preferred by the LRG sample used in this paper, while the grey, red, and blue contours assume other $\sigma_8$ determinations in the literature using $Planck$ 2018, DESI 2024, and BOSS x $Planck$ data, respectively.}
    \label{fig:kg_data}
\end{figure}

We assumed as a baseline fiducial cosmology the $\sigma_8 \sim 0.77$ measurement from the cross-correlation between ACT lensing and this LRG sample \citep{sailer2024cosmologicalconstraintscrosscorrelationdesi,kim2024atacamacosmologytelescopedr6}. As a robustness test, we recomputed the constraints assuming other $\sigma_8$ values in the literature: the $Planck$ 2018 best-fit value for this parameter, $\sigma_8 = 0.811$ \citep{1807.06209}, as well as the DESI full-shape analysis determination, $\sigma_8 = 0.842$ \citep{desicollaboration2024desi2024viicosmological} and the measurement from BOSS x $Planck$ CMB lensing, $\sigma_8 = 0.688 $ \citep{chen}. We show in Fig.~\ref{fig:kg_data} the comparison between the $C_\ell^{\kappa G}$ constraints on the $f_{\rm NL}$ - $b_0$ plane assuming the baseline and the alternative values from the literature for $\sigma_8$. Assuming the $Planck$ 2018, DESI 2024, and BOSS x LRG $\sigma_8$ best-fit values, we find $f_{\rm NL} = 42^{+44}_{-40}$, $f_{\rm NL} = 43^{+49}_{-47}$, and $f_{\rm NL} = 40 \pm 32$, respectively. The median likelihood $f_{\rm NL}$ value is unaffected in every case with respect to the $\sigma_8 = 0.77$ case listed in Tab.~\ref{tab:results}, while the 68\% confidence interval is $\sim$10\% larger for the $Planck$ 2018 case ($\sigma_8 \sim 0.81$), $\sim$20\% larger for the DESI 2024 case ($\sigma_8 \sim 0.84$), and $\sim$20\% smaller for the BOSS x LRG case ($\sigma_8 \sim 0.69$). The trend shows larger uncertainties on $f_{\rm NL}$ as a consequence of assuming larger $\sigma_8$ values. This is essentially due to the degeneracy between the galaxy bias and $\sigma_8$: higher $\sigma_8$ values result in lower $b_0$ measurements, and this affects the amplitude of the $(b-p)$ term that enhances the $f_{\rm NL}$ signal (see Eq.~\ref{deltab}).

We also tested the robustness of our results to the choice of the fiducial galaxy bias redshift evolution assumed in Eq.~\ref{eq:bz}. For this, we assumed a model in which the galaxy bias evolution was given by $b_g = b_0 \times [D(z)]^a$, and we left the $a$ parameter free, together with $f_{\rm NL}$, fixing $b_0 = 1.51$ to the best fit in Tab.~\ref{tab:results}. We find $f_{\rm NL} = 40_{-36}^{+38}$ and $a = -1.01 \pm 0.09$, obtaining full consistency with the baseline results and the fiducial galaxy bias evolution.

\begin{figure}
    \centering
\includegraphics[width=0.5\textwidth]{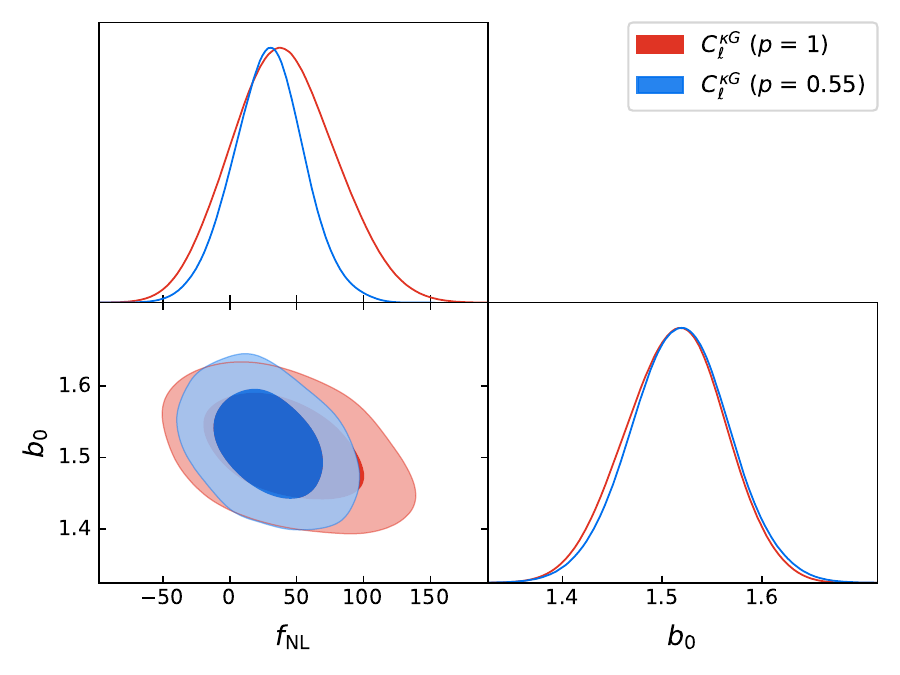}
    \caption{1$\sigma$ and 2$\sigma$ confidence ellipses for the joint posterior distribution of the $f_{\rm NL}$ and $b_0$ parameters obtained from the $C_\ell^{\kappa G}$ cross-correlation. The red contours correspond to the baseline constraints assuming $p = 1$, while the blue contours assume $p = 0.55$, as suggested in \citet{Barreira_2020}.}
    \label{fig:kg_p}
\end{figure}

Our baseline analysis assumed the universality relation for the scale-dependent galaxy bias induced by $f_{\rm NL}$, with a value $p = 1$ in Eq.~\ref{deltab}. However, works on simulations, such as \citet{Barreira_2020}, suggest a more realistic value for this parameter could be $p = 0.55$. We recomputed the constraints on $f_{\rm NL}$ from $C_\ell^{\kappa G}$, assuming $p = 0.55$, and compared them to the baseline $p=1$ case in Fig.~\ref{fig:kg_p}. When assuming $p = 0.55$, we find $f_{\rm NL} = 29_{-27}^{+25}$. The median likelihood value of $f_{\rm NL}$ is shifted as predicted by \citet{Barreira_2020}, but the compatibility with $\Lambda$CDM remains at the $\sim 1\sigma$ level. Further, we also computed the constraints for a model-independent case in which, instead of assuming the universality relation where $b_\phi = 2 \delta_c (b_g - p)$, no specific form for $b_\phi$ is assumed and we constrained the product $b_\phi f_{\rm NL}$ instead of $f_{\rm NL}$. In this case, we find $b_\phi f_{\rm NL} = 146_{-142}^{+154}$, a measurement that is still compatible with $\Lambda$CDM at the $\sim 1\sigma$ level.

\begin{figure}
    \centering
\includegraphics[width=0.5\textwidth]{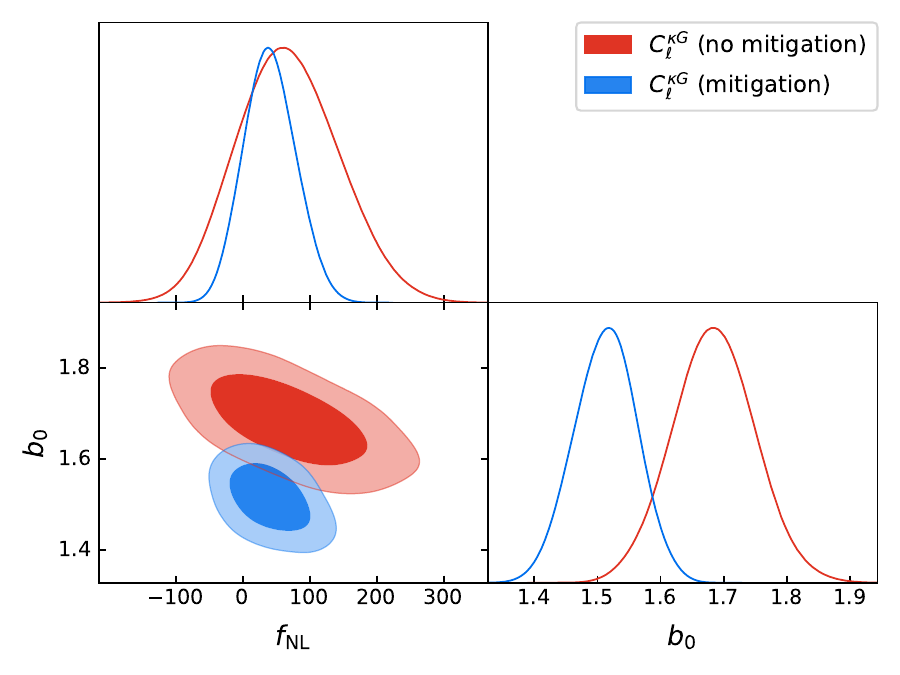}
    \caption{1$\sigma$ and 2$\sigma$ confidence ellipses for the joint posterior distribution of the $f_{\rm NL}$ and $b_0$ parameters obtained from the $C_\ell^{\kappa G}$ cross-correlation. The blue contours correspond to the baseline constraints after applying the SYSnet mitigation weights to the data and the red contours to the case in which no systematics mitigation was performed.}
    \label{fig:kg_sys}
\end{figure}

\begin{table}
   \centering
    \begin{tabular}{|c|c|c|c|c|}
    \hline
         & $C_\ell^{\kappa G}$  & $C_\ell^{GG}$ & $C_\ell^{\kappa G}$+$C_\ell^{GG}$ & $C_\ell^{\kappa G}$+$C_\ell^{GG}$ \\ \hline
        Scale cut & $\ell_{\rm min} = 2$ & $\ell_{\rm min} = 7$ & $\ell_{\rm min} = 2, 7$ & $\ell_{\rm min} = 2, 32$  \\ \hline
         $f_{\rm NL}$ & 39$^{+40}_{-38}$ & 
         $27^{+24}_{-28}$& 24$^{+20}_{-21}$   & 45$^{+38}_{-36}$ \\ \hline
         $b_0$ & 1.51$^{+0.05}_{-0.05}$& 1.48$^{+0.02}_{-0.02}$ & 1.48$^{+0.02}_{-0.02}$ & 1.48$^{+0.03}_{-0.02}$ \\ \hline
    \end{tabular}
    \caption{Median likelihood $f_{\rm NL}$ and $b_0$ values with 68\% confidence intervals obtained from the application of our analysis pipeline to the LRG data and their cross-correlation with the $Planck$ CMB lensing.}
    \label{tab:results}
    \tablefoot{The constraints from $C_\ell^{GG}$ used a scale cut at $\ell_{\rm min} = 7$ to have a better control of systematics, while the constraints from $C_\ell^{\kappa G}$ were obtained using all the multipoles up to $\ell_{\rm max} = 300$. The joint constraints also include a case in which the scale cut for $C_\ell^{GG}$ was set to $\ell_{\rm min} = 32$  in order to constrain only the galaxy bias but not $f_{\rm NL}$.}
\end{table}

\begin{figure}
    \centering
\includegraphics[width=0.5\textwidth]{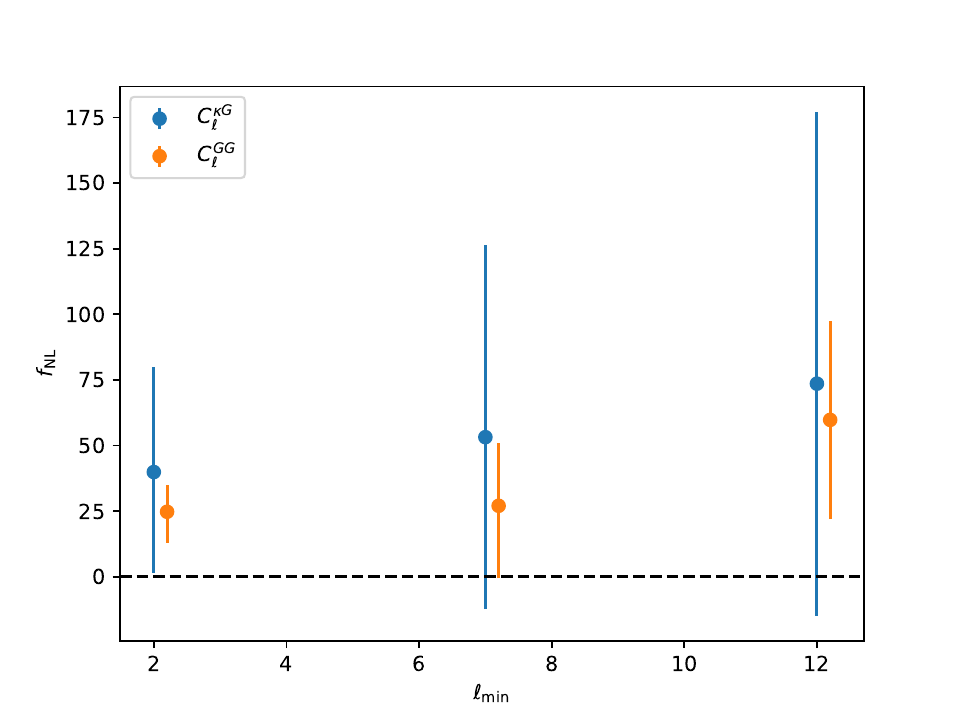}
    \caption{Constraints on $f_{\rm NL}$ from $C_\ell^{\kappa G}$ (orange dots) and $C_\ell^{GG}$ (blue dots) with 68\% confidence interval as a function of the minimum multipole $\ell_{\rm min}$ included in the analysis.}
    \label{fig:fnl_lmin}
\end{figure}

In order to understand the stability of our $f_{\rm NL}$ measurement from $C_\ell^{\kappa G}$ in terms of systematics, we compare in Fig.~\ref{fig:kg_sys} the constraints from the $Planck$ lensing - DESI LRG cross-correlation with the case in which no systematics mitigation was performed. In the latter case, we find $f_{\rm NL} = 64_{-73}^{+80}$. The result is compatible with the baseline case using systematics weights, but the uncertainties are larger due to the impact of the extra $C_\ell^{G G}$ power spectrum in the covariance matrix. This also emphasizes the importance of performing an accurate systematics mitigation, even for cross-correlation analysis.

We also tested the robustness of our results to the systematics looking at the stability of the results as a function of the minimum multipole, $\ell_{\rm min}$. In Fig.~\ref{fig:fnl_lmin} we represent the constraints on $f_{\rm NL}$ from $C_\ell^{\kappa G}$ and $C_\ell^{GG}$ for $\ell_{\rm min}$ = 2, 7, and 12. For $C_\ell^{\kappa G}$, the constraints are compatible at a $\lesssim 1 \sigma$ level with $f_{\rm NL} = 0$ for the three $\ell_{\rm min}$ values. Instead, for $C_\ell^{GG}$, we find a $\sim 1 \sigma$ deviation for $\ell_{\rm min} = 7$, while this deviation rises to $\sim 2 \sigma$ for $\ell_{\rm min} = 2$ and $\ell_{\rm min} = 12$. This indicates that the lower multipoles on $C_{\ell}^{GG}$ might be dominated by observational systematics, as suggested by the test on mocks, while for $C_\ell^{\kappa G}$ it is safer to include all the scales in the analysis.

\begin{table}
    \centering
    \begin{tabular}{|c|c|c|c|}
        \hline
        Region  & $C_\ell^{\kappa G} (\ell_{\rm min} = 2)$  & $C_\ell^{GG} (\ell_{\rm min} = 7)$ & $C_\ell^{\kappa G}$+$C_\ell^{GG}$  \\ \hline
        NGC & 90$^{+208}_{-174}$ & 30 $\pm$ 43 
          &  20 $\pm$ 40   \\ \hline
         SGC & 82$^{+118}_{-112}$  & 26$^{+28}_{-31}$  & 36$^{+26}_{-29}$ \\ \hline
    \end{tabular}
    \caption{Consistency check for the median likelihood $f_{\rm NL}$ values with 68\% confidence intervals obtained from the application of our analysis pipeline to the different sky regions (NGC and SGC) of the LRG data and their cross-correlation with the $Planck$ CMB lensing.}
    \label{tab:regions}
\end{table}

In addition, we tested the compatibility of our results among different regions of the sky. We computed our 
constraints on $f_{\rm NL}$ by applying masks of the NGC and SGC and list the results in Tab.~\ref{tab:regions}. We find compatible results between the two galactic caps, and even if the best-fit $f_{\rm NL}$ values we obtain from $C_\ell^{\kappa G}$ are consistently higher than those measured from the full footprint, the measurements are compatible with zero PNG, considering the larger error bars.

Our constraint on $f_{\rm NL}$ from the DESI LRG - $Planck$ lensing cross-correlation is also compatible with the main result from \cite{rezaie2023local} using the LRG autocorrelation only. \cite{rezaie2023local} discuss in detail the impact of the mitigation recipe on the $f_{\rm NL}$ error bars and best-fit values. They highlight $f_{\rm NL} = 34_{-44}^{+24}$ when applying a more aggressive nonlinear regression for systematics mitigation using nine features maps, while we selected a more conservative treatment in order to avoid a strong overfit in the cross-correlation, $C_{\ell}^{\kappa G}$. When applying the same nonlinear treatment with three maps that we chose for this study, they find $f_{\rm NL} = 28_{-11}^{+12}$ (before calibration) and $f_{\rm NL} = 47 \pm 14$ (after calibration). We note that, in \cite{rezaie2023local}, a calibration of the constraints is proposed, based on the results of the tests on mocks, and that all multipoles were included in the baseline analysis, while we did not apply any calibration and we set $\ell_{\rm min} = 7$ for $C_{\ell}^{GG}$. 
Another important difference in our analysis is the contamination model on the mocks: we used nonlinear weights to implement the contamination on the mocks, while in \cite{rezaie2023local}, linear weights were applied to generate this contamination. A calibration of the $f_{\rm NL}$ constraints based on the mocks would be model dependent in any case, and our aim is instead to show the capabilities of $C_{\ell}^{\kappa G}$ to produce more stable measurements of $f_{\rm NL}$, either alone or in combination with $C_{\ell}^{GG}$.

\section{Conclusions}
\label{sec:conclusions}

In this work, we used an LRG catalog from the DESI Legacy imaging surveys, calibrated with the spectroscopic redshifts that have been observed for the DESI Survey Validation, in combination with the $Planck$ PR4 CMB lensing map, to put a constraint on the primordial local non-Gaussianity parameter, $f_{\rm NL}$. We measured $f_{\rm NL}$ through the scale-dependent bias effect using as observables the cross-correlation between the LRG and CMB lensing maps in the angular domain, $C_\ell^{\kappa G}$, and the autocorrelation of the LRG field, $C_\ell^{GG}$.

In order to limit the impact of imaging systematics on large scales where the $f_{\rm NL}$ signal is present, we used a neural network code for imaging systematics mitigation. Our measurement was performed without blinding, but the full analysis methodology was tested on mock fields including imaging systematics for different $f_{\rm NL}$ values. Our end-to-end pipeline works at a reasonable level of agreement with the input $f_{\rm NL}$ values, being especially robust when combining both $C_\ell^{\kappa G}$ and $C_\ell^{GG}$ for the positive and zero $f_{\rm NL}$ mocks. From the systematics tests on mocks, we find residual contamination in the behavior of the five first multipoles ($\ell = 2-6$) of $C_\ell^{GG}$; hence, we imposed a cut at $\ell_{\rm min} = 7$ for the autocorrelation power spectrum. 

From the CMB - LRG cross-correlation, $C_\ell^{\kappa G}$, alone we find $f_{\rm NL} = 39_{-38}^{+40}$ at the 68\% confidence level. We performed some robustness tests on this result, such as changing the fiducial $\sigma_8$ to other values in the literature, different from the DESI LRG best fit, changing the value of the $p$ parameter, modifying the fiducial bias evolution, or varying the minimum multipole, $\ell_{\rm min}$, to evaluate the possible impact of systematics, and we find consistent results for every test. If we combine our result with the information from the angular LRG autocorrelation, $C_\ell^{GG}$, adopting the $\ell_{\rm min} = 7$ cut, we find $f_{\rm NL} = 24_{-21}^{+20}$, although $C_\ell^{GG}$ presents a larger statistical fluctuation as a function of $\ell_{\rm min}$, suggesting that $C_\ell^{\kappa G}$ is more stable and less sensitive to the effect of imaging systematics. Our results are consistent with the $f_{\rm NL}$ measurements from the DESI LRG by \cite{rezaie2023local} and motivate the future use of CMB cross-correlations, together with the tracers' autocorrelations and an appropriate systematics mitigation technique, in order to perform accurate $f_{\rm NL}$ measurements. 

\section*{Data availability}
All data points, maps, covariances, and MCMC chains shown in the figures of this paper are publicly available in \url{https://doi.org/10.5281/zenodo.14401463.}

\begin{acknowledgements}
J. R. Bermejo-Climent and R. Demina acknowledge support from the U.S. Department of Energy under the grant DE-SC0008475.0. JRBC acknowledges the support of the Spanish Ministry of Science and Innovation under the grants PID2021-126616NB-I00 and ``DarkMaps'' PID2022-142142NB-I00, and from the European Union through the grant ``UNDARK'' of the Widening participation and spreading excellence program (project number 101159929). AK was supported as a CITA National Fellow by the Natural Sciences and Engineering Research Council of Canada (NSERC), funding reference \#DIS-2022-568580. MR is supported by the U.S. Department of Energy grants DE-SC0021165 and DE-SC0011840.

This material is based upon work supported by the U.S. Department of Energy (DOE), Office of Science, Office of High-Energy Physics, under Contract No. DE–AC02–05CH11231, and by the National Energy Research Scientific Computing Center, a DOE Office of Science User Facility under the same contract.

Additional support for DESI was provided by the U.S. National Science Foundation (NSF), Division of Astronomical Sciences under Contract No. AST-0950945 to the NSF’s National Optical-Infrared Astronomy Research Laboratory; the Science and Technology Facilities Council of the United Kingdom; the Gordon and Betty Moore Foundation; the Heising-Simons Foundation; the French Alternative Energies and Atomic Energy Commission (CEA); the National Council of Humanities, Science and Technology of Mexico (CONAHCYT); the Ministry of Science, Innovation and Universities of Spain (MICIU/AEI/10.13039/501100011033), and by the DESI Member Institutions: \url{https://www.desi.lbl.gov/collaborating-institutions}.

The DESI Legacy Imaging Surveys consist of three individual and complementary projects: the Dark Energy Camera Legacy Survey (DECaLS), the Beijing-Arizona Sky Survey (BASS), and the Mayall z-band Legacy Survey (MzLS). DECaLS, BASS and MzLS together include data obtained, respectively, at the Blanco telescope, Cerro Tololo Inter-American Observatory, NSF’s NOIRLab; the Bok telescope, Steward Observatory, University of Arizona; and the Mayall telescope, Kitt Peak National Observatory, NOIRLab. NOIRLab is operated by the Association of Universities for Research in Astronomy (AURA) under a cooperative agreement with the National Science Foundation. Pipeline processing and analyses of the data were supported by NOIRLab and the Lawrence Berkeley National Laboratory. Legacy Surveys also uses data products from the Near-Earth Object Wide-field Infrared Survey Explorer (NEOWISE), a project of the Jet Propulsion Laboratory/California Institute of Technology, funded by the National Aeronautics and Space Administration. Legacy Surveys was supported by: the Director, Office of Science, Office of High Energy Physics of the U.S. Department of Energy; the National Energy Research Scientific Computing Center, a DOE Office of Science User Facility; the U.S. National Science Foundation, Division of Astronomical Sciences; the National Astronomical Observatories of China, the Chinese Academy of Sciences and the Chinese National Natural Science Foundation. LBNL is managed by the Regents of the University of California under contract to the U.S. Department of Energy. The complete acknowledgments can be found at \url{https://www.legacysurvey.org/}.

Any opinions, findings, and conclusions or recommendations expressed in this material are those of the author(s) and do not necessarily reflect the views of the U. S. National Science Foundation, the U. S. Department of Energy, or any of the listed funding agencies.

The authors are honored to be permitted to conduct scientific research on Iolkam Du’ag (Kitt Peak), a mountain with particular significance to the Tohono O’odham Nation.
\end{acknowledgements}

\bibliographystyle{aa} 
\bibliography{aanda.bib} 
\end{document}